\documentclass[]{aastex63}
\usepackage{graphicx} 
\usepackage{amsmath} 
\usepackage{amssymb} 
\usepackage{cases}
\usepackage{mathrsfs} 
\usepackage[flushleft]{threeparttable}
\usepackage{amssymb}
\usepackage{amsmath}
\usepackage{epstopdf}
\usepackage{natbib}
\usepackage{xcolor}
\usepackage[all]{hypcap}
\usepackage[flushleft]{threeparttable}
\usepackage{mwe,tikz}
\usepackage{bm}

\newcommand{\e}{\text{e}}

\newcommand{\p}{\partial}

\accepted{June, 2020}
\submitjournal{ApJ}

\shorttitle{Implications of GBM-190816}
\shortauthors{Yang et al.}
\begin{document}

\title{Physical Implications of the Sub-threshold GRB GBM-190816 and its Associated Sub-threshold Gravitational Wave Event}
\correspondingauthor{Bin-Bin Zhang, Bing Zhang}
\email{bbzhang@nju.edu.cn, zhang@physics.unlv.edu}

\author[0000-0002-7555-0790]{Yi-Si Yang}
\affiliation{School of Astronomy and Space Science, Nanjing
University, Nanjing 210093, China}
\affiliation{Key Laboratory of Modern Astronomy and Astrophysics (Nanjing University), Ministry of Education, China}

\author[0000-0002-1766-6947]{Shu-Qing Zhong}
\affiliation{School of Astronomy and Space Science, Nanjing
University, Nanjing 210093, China}
\affiliation{Key Laboratory of Modern Astronomy and Astrophysics (Nanjing University), Ministry of Education, China}
\author[0000-0003-4111-5958]{Bin-Bin Zhang}
\affiliation{School of Astronomy and Space Science, Nanjing
University, Nanjing 210093, China}
\affiliation{Key Laboratory of Modern Astronomy and Astrophysics (Nanjing University), Ministry of Education, China}
\affiliation{Department of Physics and Astronomy, University of Nevada Las Vegas, NV 89154, USA}

\author[0000-0002-9188-5435]{Shichao Wu}
\affiliation{Department of Astronomy, Beijing Normal University, Beijing 100875, China}

\author[0000-0002-9725-2524]{Bing Zhang}
\affiliation{Department of Physics and Astronomy, University of Nevada Las Vegas, NV 89154, USA}

\author[0000-0003-0691-6688]{Yu-Han Yang}
\affiliation{School of Astronomy and Space Science, Nanjing
University, Nanjing 210093, China}
\affiliation{Key Laboratory of Modern Astronomy and Astrophysics (Nanjing University), Ministry of Education, China}

\author[0000-0002-1932-7295]{Zhoujian Cao}
\affiliation{Department of Astronomy, Beijing Normal University, Beijing 100875, China}

\author[0000-0002-3100-6558]{He Gao}
\affiliation{Department of Astronomy, Beijing Normal University, Beijing 100875, China}

\author{Jin-Hang Zou}
\affiliation{Department of Space Sciences and Technology, Hebei Normal University, Shijiazhuang 050024, China}

\author[0000-0002-2662-6912]{Jie-Shuang Wang}
\affiliation{Tsung-Dao Lee Institute, Shanghai Jiao Tong University, Shanghai 200240, China}

\author[0000-0001-6396-9386]{Hou-Jun L\"{u}}
\affiliation{Guangxi Key Laboratory for Relativistic Astrophysics, Department of Physics, Guangxi University, Nanning 530004, China}

\author{Ji-Rong Cang}
\affiliation{Department of Astronomy, Tsinghua University, Beijing 100084, China }

\author[0000-0002-7835-8585]{Zi-Gao Dai}
\affiliation{School of Astronomy and Space Science, Nanjing
University, Nanjing 210093, China}
\affiliation{Key Laboratory of Modern Astronomy and Astrophysics (Nanjing University), Ministry of Education, China}

\begin{abstract}

The LIGO-Virgo and Fermi collaborations recently reported a possible joint detection of a sub-threshold gravitational wave (GW) event and a sub-threshold gamma-ray burst (GRB), GBM-190816, that occurred 1.57 s after the merger. We perform an independent analysis of the publicly available data and investigate the physical implications of this potential association. By carefully studying the following properties of GBM-190816 using {\it Fermi}/GBM data, including signal-to-noise ratio, duration, \textit{f}-parameter, spectral properties, energetic properties, and its compliance with some GRB statistical correlations, we confirm that this event is likely a typical short GRB. Assuming its association with the sub-threshold GW event, the inferred luminosity is $1.47_{-1.04}^{+3.40} \times 10^{49}$ erg s$^{-1}$. Based on the available information of the sub-threshold GW event, we infer the mass ratio $q$ of the compact binary as $q=2.26_{-1.43}^{+2.75}$ (90\% confidence interval) according to the reported range of luminosity distance. If the heavier compact object has a mass $>$ 3 solar masses, $q$ can be further constrained to $q=2.26_{-0.12}^{+2.75}$. The leading physical scenario invokes an NS-BH merger system with the NS tidally disrupted. Within this scenario, we constrain the  physical properties of such a system (including mass ratio $q$, the spin parameters, and the observer's viewing angle) to produce a GRB. The GW data may also allow an NS-BH system with no tidal disruption of the NS (the plunge events) or a BH-BH merger. We apply the charged compact binary coalescence (cCBC) theory (for both a constant charge and an increasing charge for the merging members) to derive the model parameters to account for GBM-190816 and found that the required parameters are extreme. Finally, we argue that the fact that the observed GW-GRB delay time scale is comparable to that of GW170817/GRB 170817A suggests that the GW-GRB time delay of these two cases is mainly defined by the time scale for the jet to propagate to the energy dissipation / GRB emission site.

\end{abstract}


\section{Introduction} \label{sec:intro}

The field of gravitational wave (GW)-led multi-messenger astrophysics grows rapidly since the detection of the first gravitational wave (GW) event from a binary black hole (BH-BH) merger GW150914 \citep{Abbott2016}, and especially after the detection of the first GW event from a binary neutron star (NS-NS) merger system that was associated with electromagnetic (EM) signals, GW170817/GRB 170817A \citep{Abbott2017,Goldstein2017,Savchenko2017}. Searching for EM counterparts coincident with gravitational wave signals from different types of compact binary mergers has been a long-sought goal in the field. Since the start of the LIGO O3 observational run, many follow-up observations of GW events using space-borne or ground-based multi-messenger facilities have been carried out, but so far no high-confidence detection has been made.

One interesting event was a sub-threshold GRB candidate, Fermi GBM-190816, which was potentially associated with a subthreshold LIGO/Virgo compact binary merger candidate, as reported by the LIGO/Virgo/Fermi collaborations in \cite{Goldstein2019} and \cite{Goldstein2019a}. The gamma-ray signal was registered by $Fermi$/GBM \citep{Meegan2009} at 21:22:14.563 16th August 2019 UTC (hereafter $T_0$), which was about 1.57 s after a possible sub-threshold gravitational wave (GW) signal detected by LIGO/Virgo \citep{Goldstein2019}. The GW signal, proposed to be a possible compact binary coalescence (CBC) candidate, is located at a distance\footnote{The distance information can be found from the header of the fits file in \url{https://wiki.gw-astronomy.org/pub/OpenLVEM/FermiGBM-LVC/GBM-190816-with-LV-v2.fits}. We acknowledge Eric Burns for pointing this out to us.} \citep{Shawhan2019} of $428 \pm 143$ Mpc (90\% confidence interval [CI]), about nine times farther than the distance of GW170817/GRB 170817A. According to the GW signal, the lighter compact object is estimated to be lighter than three solar masses, which can be either an NS or a low-mass BH that merges with a higher-mass BH. 

Since it takes long for the LIGO/Virgo/Fermi collaborations to release the official results, we decided to independently process the publicly available data and investigate the physical implications of such a putative association. We first perform a detailed analysis of the sub-threshold gamma-ray signal observed by {\it Fermi}/GBM to confirm its consistency with a short GRB (\S 2). Based on the available information about the GW event (e.g. the fact that the GW signal is sub-threshold and the lighter member has mass $<3 M_\odot$), we then estimate the mass ratio $q$ of the binary system (\S 3). Next, using the observed EM properties, we constrain the physical properties of the system for several astrophysical scenarios, including NS-BH mergers with and without tidal disruption as well as BH-BH mergers (\S 4). The physical implications of the 1.57 s GW-GRB delay are also discussed in \S 4. Our results are summarized in \S 5.

\section{The Sub-threshold Burst}
\subsection{Data Reduction and Selection}

\begin{figure*}
 \label{fig:14detlc}
 \centering
 \includegraphics[width=1.0\textwidth]{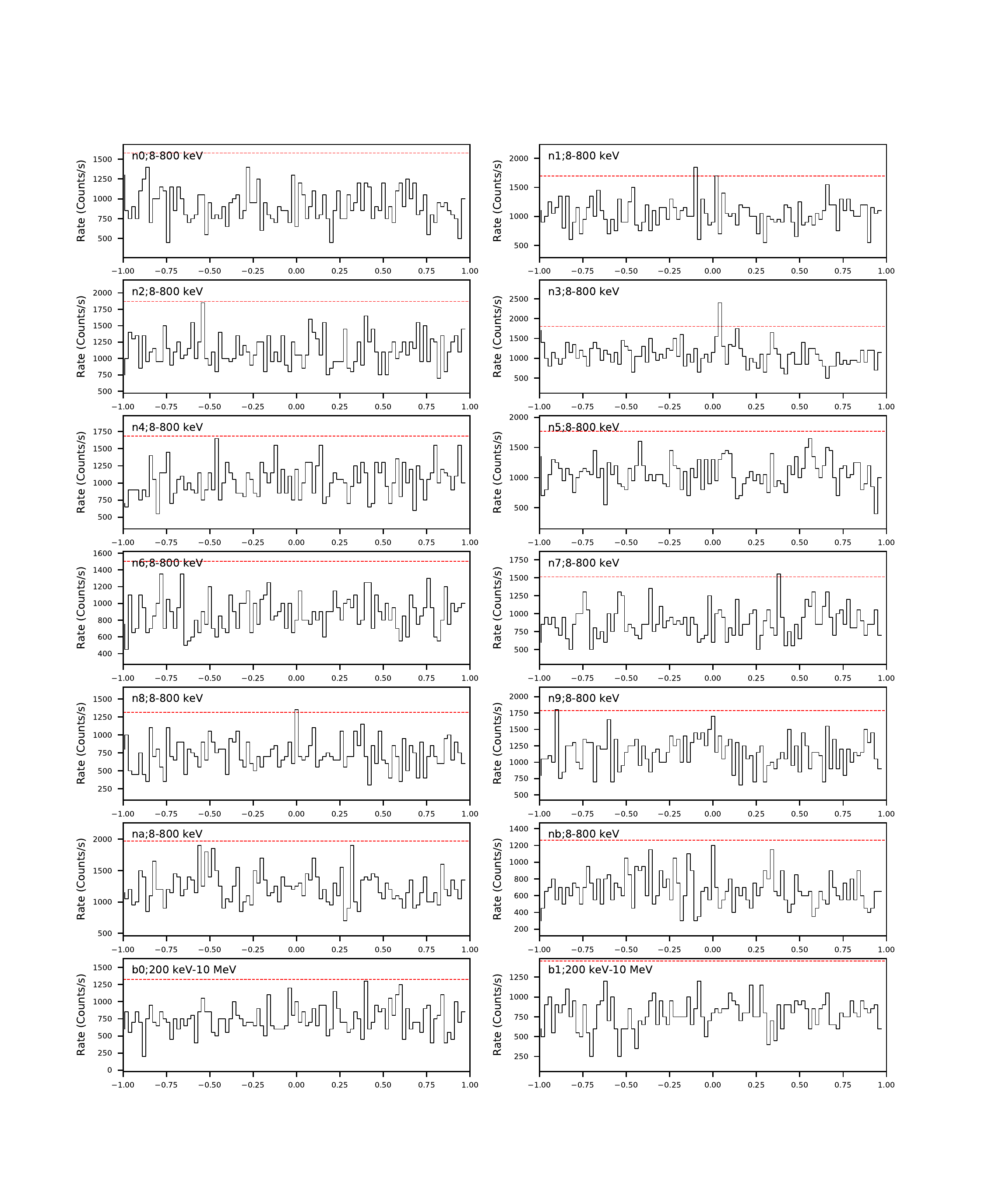} 
 \caption{Light curves around $T_0$ of all 14 detectors. The time bin size is 0.02 s. N3 detector panel shows a sharp peak around $T_0$. The red horizontal dashed lines represent the 3$\sigma$ level for each detector. }
\end{figure*}

We download the corresponding Time-Tagged-Event data from the public data site of {\em Fermi}/GBM according to the time of the event reported by \cite{Goldstein2019}. Data reduction follows the standard procedure, as discussed in \cite{BBZhang2011,BBZhang2016,BBZhang2018}. The full-energy-range light curves of all fourteen GBM detectors are shown in Figure \ref{fig:14detlc}. The weak sub-threshold GRB is visible in the light curve of the NaI detector n3 and marginally visible in n1. Indeed, using the best-fit location (178.23$^{\circ}$, 33.52$^{\circ}$) of the GW signal, we calculate that NaI detectors n1 \& n3 hold the smallest angular separations with respect to the GW source. Thus, those two detectors are selected for further temporal and spectral analysis. No BGO detector is selected as no significant emission has been observed above 800 keV.

\subsection{Burst Properties}

\begin{figure*}
 \centering
 \includegraphics[width=0.6\textwidth]{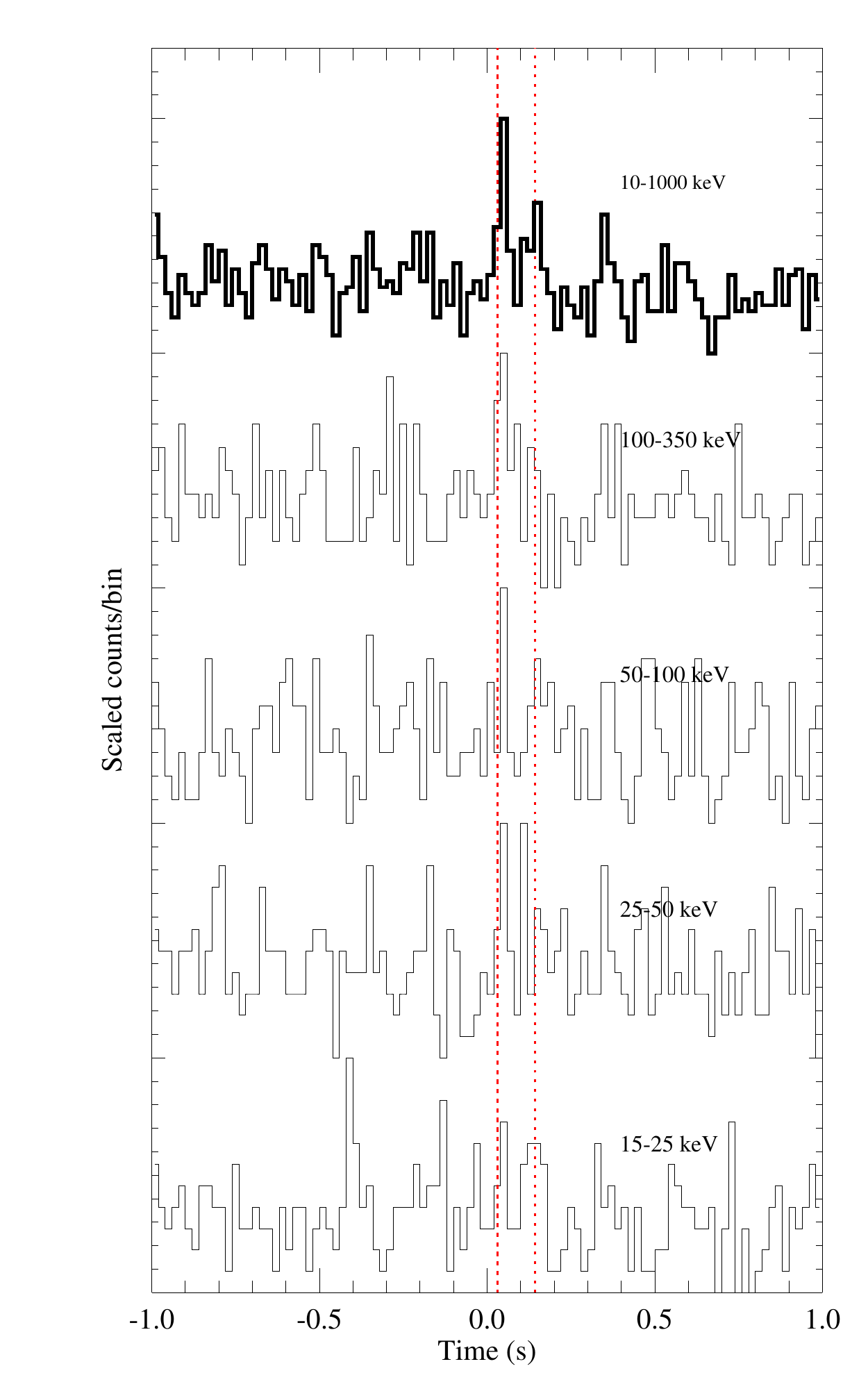} 
 \caption{Energy dependent light curves of detector n3. Vertical dashed lines mark the $T_{90}$ interval.}
 \label{fig:Lightcurve}
\end{figure*}

We perform the following analysis on the gamma-ray signal \citep{BBZhang2011,BBZhang2016,BBZhang2018} to study the properties of GBM-190816:

(1) Signal confirmation. We analyze the TTE data of the detector n3 using the Bayesian Block (BB) algorithm \citep{Scargle2013}. Searching in the interval from $T_0 - 10$ s to $T_0 + 10$ s, we find a significant sharp signal starting from $T_0 + 0.038$ s to $T_0 + 0.056$ s. We then try to derive the significance level of the burst. The background is taken from two intervals $T_0 - 15$ s to $T_0 - 5$ s and $T_0 + 5$ s to $T_0 + 15$ s. By varying the energy band and the bin size (we make sure that there are at least two bins in the burst block), we find the signal-to-noise ratio (S/N) reaching 3.95. Figure \ref{fig:Lightcurve} shows the light curves in four different energy channels for detector n3. The details of this method can be found in \cite{Wang2019}. The False Alarm Rate (FAR) of detecting such an event is about $1.2\times10^{-4}$ \citep{Goldstein2019}.

\begin{figure*}
	\centering
	\includegraphics[width=0.5\linewidth]{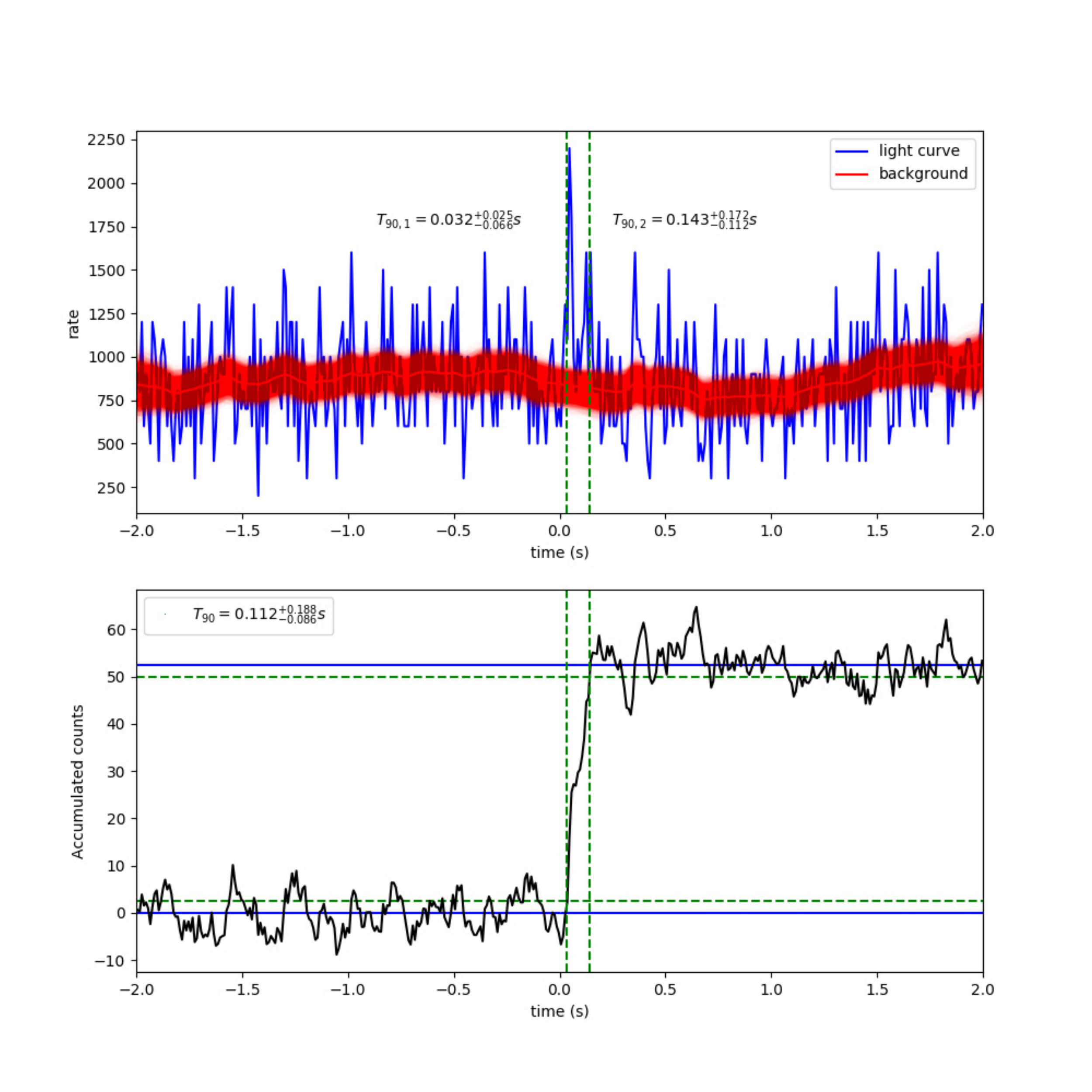}
	\caption{T90 calculation. Upper panel: The blue curve is the light curve plotted with the n3 data. The red curve represents background baseline fitted by the MCMC method. Shaded red regions mark the 1$\sigma$ region. The $T_{90}$ intervals of GBM-190816, $T_0 + 0.032_{-0.065}^{+0.025}$ s and $T_0 + 0.143_{-0.11}^{+0.17}$ s are marked with the green dashed lines. Lower panel: Accumulated light curve. Blue horizontal lines are average levels of accumulated counts, green horizontal dashed lines represent 5\% and 95\% of accumulated counts, which are used to calculate $T_{90}$.}
	\label{fig:t90}
\end{figure*}

(2) Burst duration. For simplicity, we estimate $T_{90}$ of the burst based on the cumulative net count rate. The background is estimated by applying the ``baseline" method \citep{BBZhang2018} to some long time intervals before and after the signal region. By calculating the time interval during which $90\%$ of the total net counts have been detected, we obtained $T_{90}= 0.112_{-0.085}^{+0.185}$ s with the starting and ending time $T_{90,1}= 0.032_{-0.065}^{+0.025}$ s and $T_{90,2}= 0.143_{-0.11}^{+0.17}$ s, respectively (Figure \ref{fig:t90}). The uncertainties are calculated by a Monte Carlo approach, which takes into account the fluctuations of the observed light curve. 

\begin{figure*}
\begin{tabular}{lr}
\includegraphics[angle=0,width=0.47\textwidth]{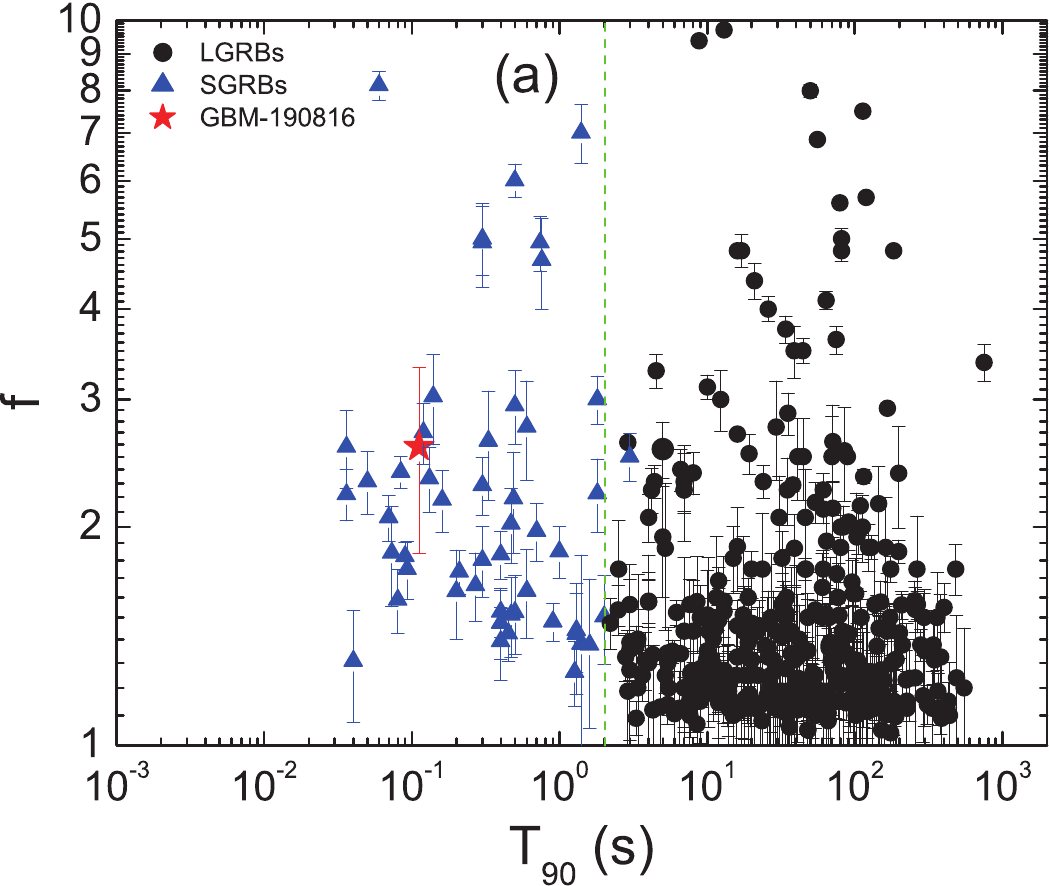} &
\includegraphics[angle=0,width=0.5\textwidth]{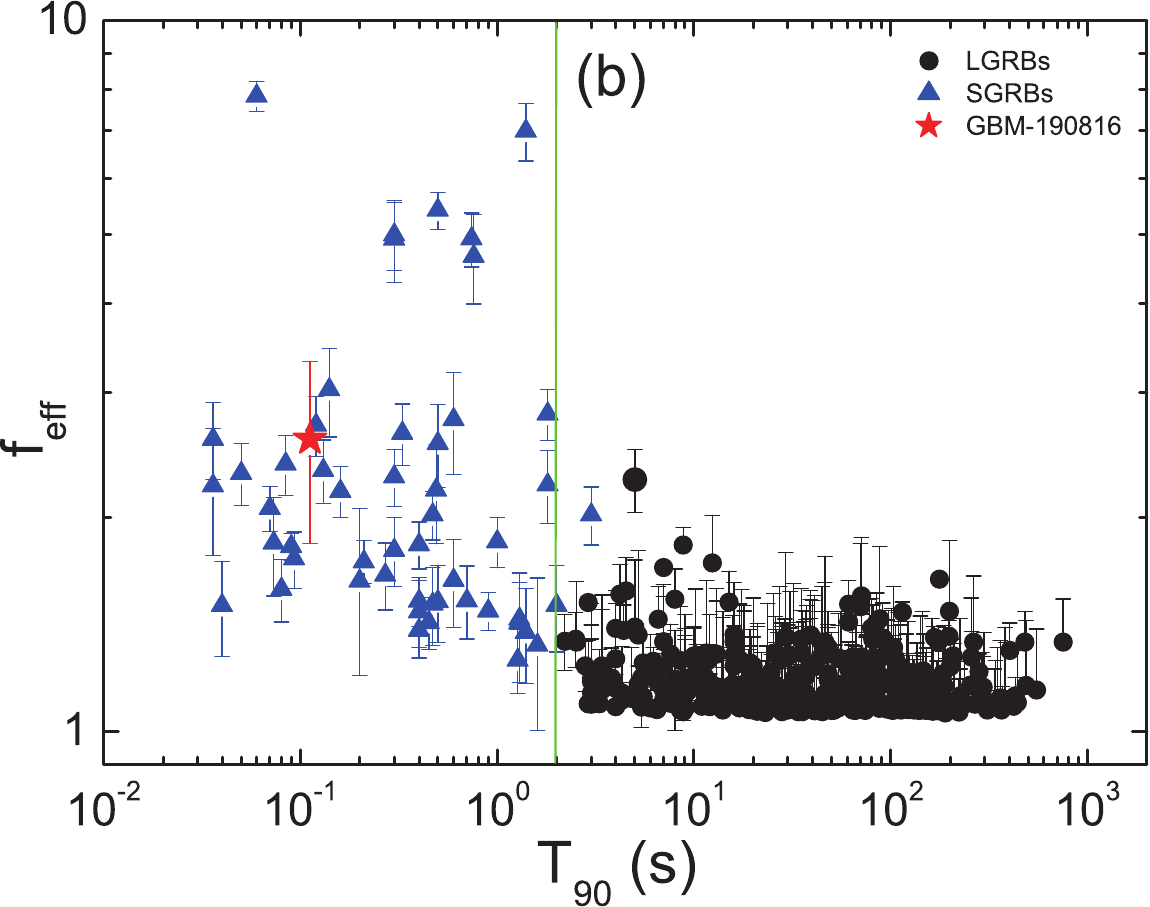}
\end{tabular}

\caption{The $f$ and $f_{\rm eff}$ parameters of GBM-190816 and their comparisons with other short and long GRBs.}
\label{fig:fpara}

\end{figure*}

(3) Amplitude parameter of GBM-190816. \cite{lv2014} defined two ``amplitude parameters'' to assist burst classifications: the parameter $f$ denotes the ratio between the peak flux and the average background flux, and $f_{\rm eff}$ denotes the ratio between the peak flux of a pseudo-burst and the average background flux. The pseudo-burst is defined by scaling down the peak flux until the measured duration of a long burst is shorter than two seconds \citep{lv2014}. For short GRBs, $f=f_{\rm eff}$. Statistically, the $f_{\rm eff}$ parameters of long GRBs are typically smaller than $f$ of short GRBs, providing a criterion to identify contaminated long GRBs in the observed short GRB sample due to the ``tip-of-iceberg'' effect. We  perform the $f$ analysis to GBM-190816, and obtain $f=2.58\pm0.37$. Figure \ref{fig:fpara}(a) and (b) show $T_{90}$ as functions of $f$ and $f_{\rm eff}$ for both long and short GRBs, where GBM-190816 is highlighted as a star. We find that its amplitude parameter is generally larger than $f_{\rm eff}$ of typical long GRBs, consistent with being a typical short GRB. Moreover, we calculate the probability of GBM-190816 being a disguised short GRB according to the $p-f$ relation derived by \cite{lv2014}. We find that such a probability is $p\sim0.03$. All these suggest that GBM-190816 is a genuine short GRB. Nonetheless, there is a non-negligible probability that the observed spike could be still the ``tip of iceberg" of a longer short burst (see more discussion in \S\ref{sec:timedelay}).

\begin{figure*}
\begin{tabular}{lll}
\includegraphics[angle=0,width=0.33\textwidth]{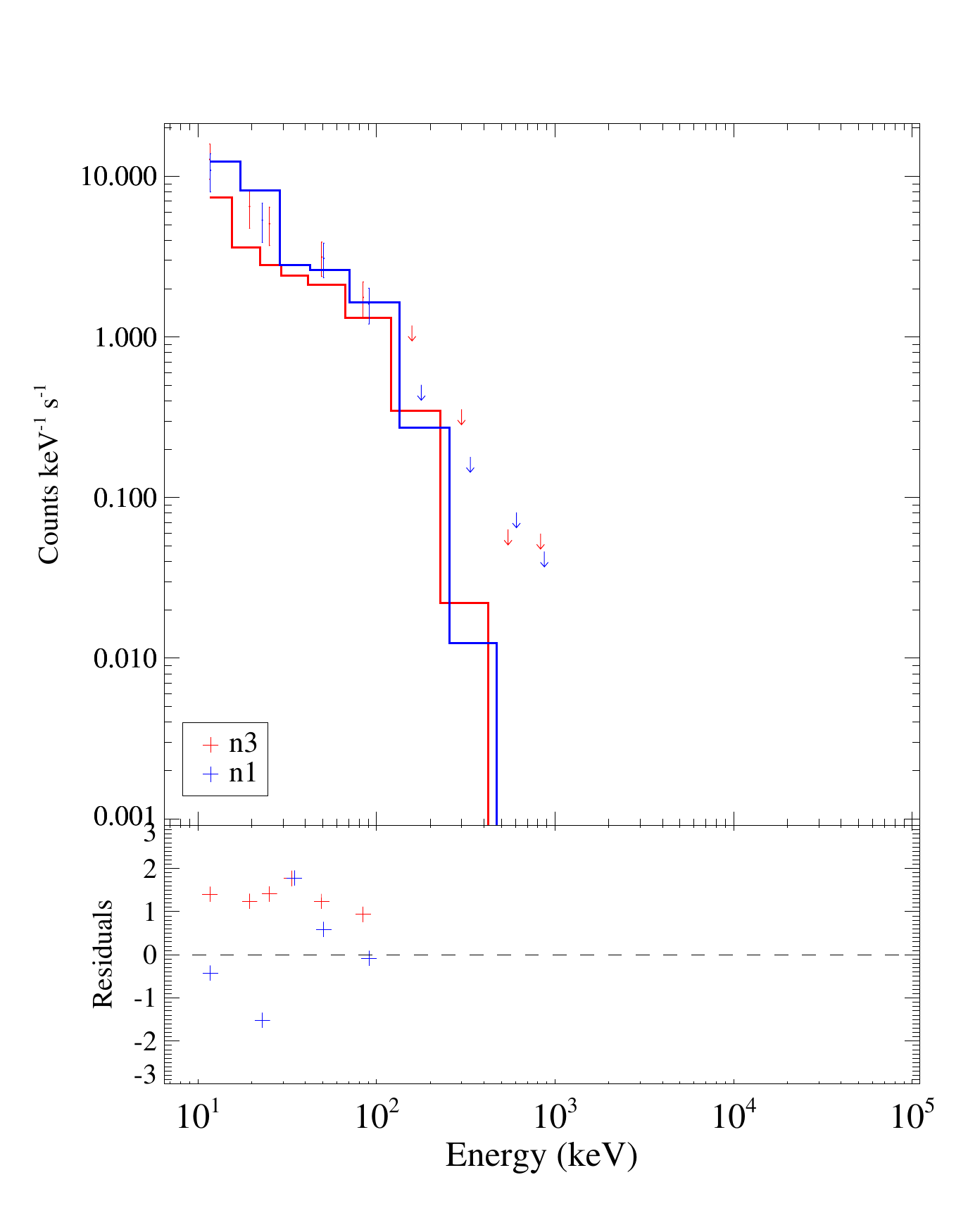}
\includegraphics[angle=0,width=0.33\textwidth]{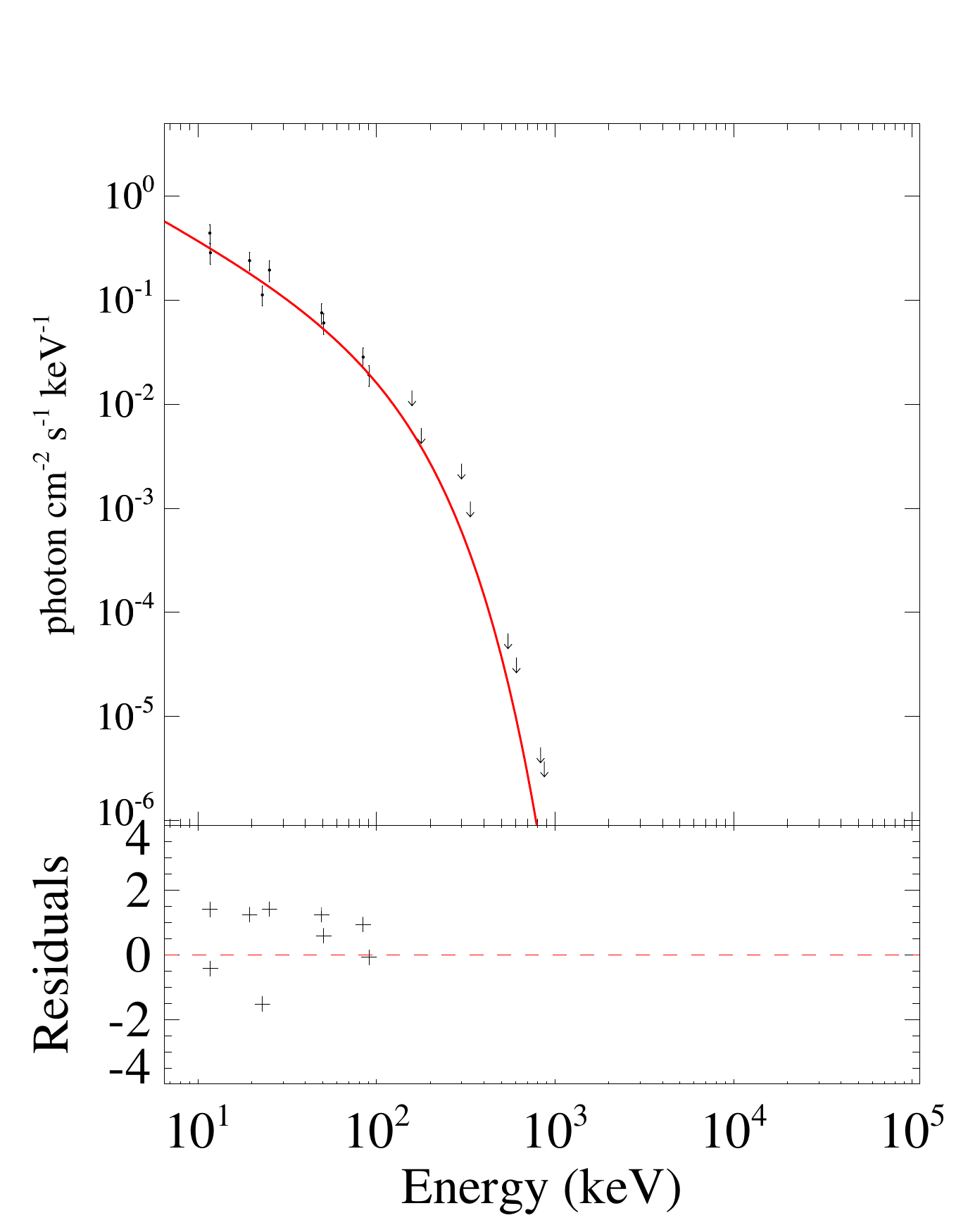}
\includegraphics[angle=0,width=0.33\textwidth]{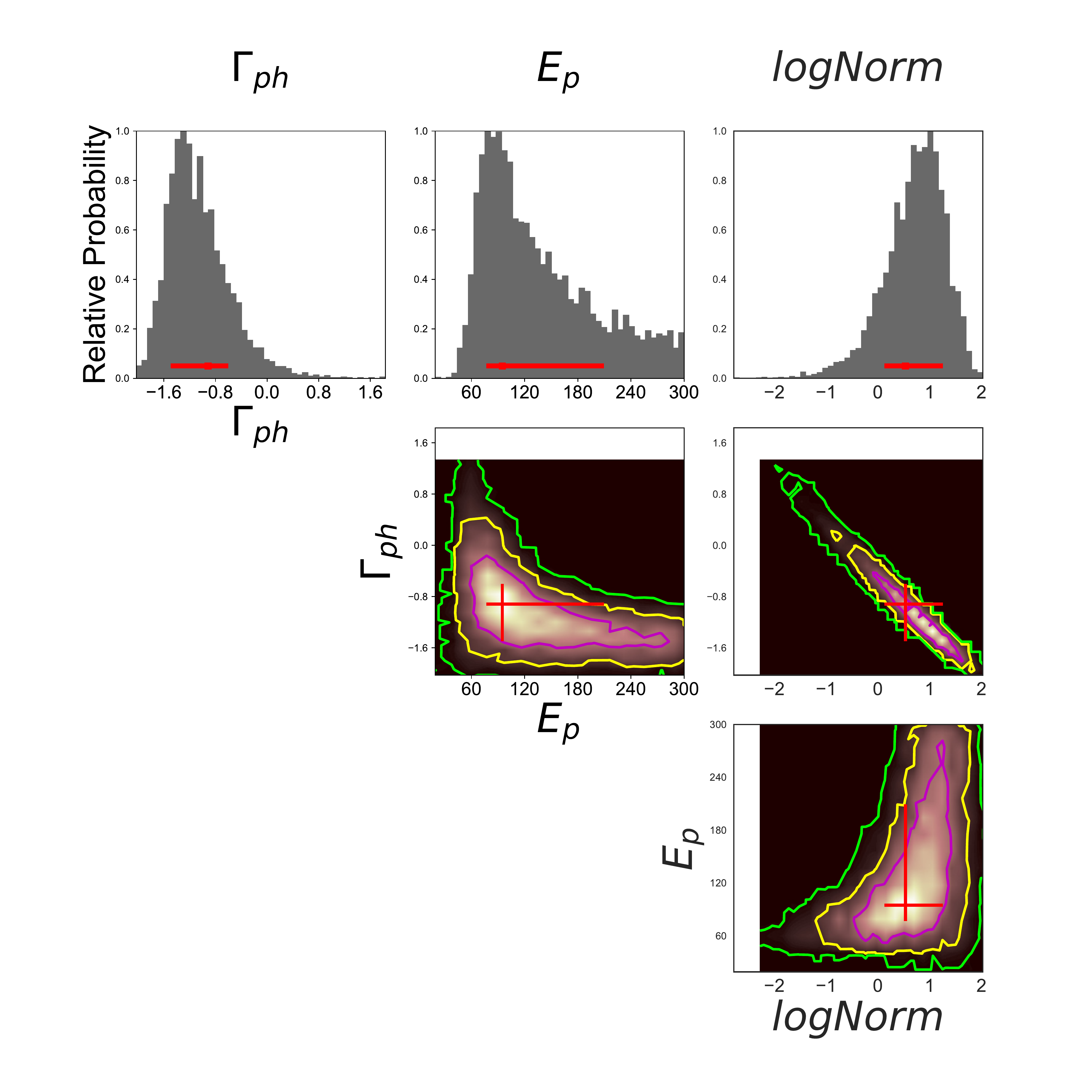}
\end{tabular}

\caption{\textit{Left}: The observed count spectrum of GBM-190816 within the $T_{90}$ time interval and its fit by the CPL model. \textit{Middle}: De-convolved photon spectrum. \textit{Right}: Parameter constraints of the CPL fit. Histograms and contours in the corner plots illustrate the likelihood 2-D map. Red crosses show the best-fitting values. All error bars in these panels represent the 1$\sigma$ uncertainties. }
\label{fig:corner}

\end{figure*}

(4) Spectral analysis. 
We extract the time-integrated spectra of GBM-198016 between $T_{90,1}$ and $T_{90,2}$. Only two GBM detectors, n1 \& n3, are selected due to the reasons mentioned in \S 2.1. Background spectra are obtained by empirically modeling the source-free time intervals around the burst. The detector response matrices (DRMs), which are needed in the spectral fitting is generated using the response generator provided by the {\it Fermi} Science Tools\footnote{\url{https://fermi.gsfc.nasa.gov/ssc/data/analysis/rmfit/gbmrsp-2.0.10.tar.bz2}}. Spectral fitting is performed using $McSpecfit$ \citep{BBZhang2018}. A handful of spectral models, such as simple power-law (PL), cutoff power-law (CPL), Band function (Band), Blackbody (BB), and the combinations of any two or three models, are considered to fit the observed spectra. We then compare the goodness of the fits and find that the CPL is the best one that adequately describes the observed data according to the Bayesian Information Criteria (BIC). The CPL model fit (Figure \ref{fig:corner}) gives a peak energy of $94.84_{-17.94}^{+114.64}$ keV and a lower energy spectral index of $-0.92_{-0.58}^{+0.32}$, both being typical for GRB spectral parameters. The best-fit parameters of CPL fits are listed in Table 1. No further time-resolved spectral fitting is performed due to the low number of photon counts.

(5) Burst energy. Using the the best-fit parameters of the CPL model, we find that the average flux within $T_{90}$ is $6.65_{-2.26}^{+5.72} \times 10^{-7} \rm {erg\,cm^{-2}\,s^{-1}}$ between 1 keV to 10000 keV. The total fluence in the same energy range is $7.38_{-2.51}^{+6.35} \times 10^{-8} \rm {erg\,cm^{-2}}$. Taking into account the burst distance $\sim 428 $ Mpc, we further calculate the corresponding isotropic luminosity and energy as $L_{\gamma,\rm iso}= 1.47_{-1.04}^{+3.40} \times 10^{49}$ erg s$^{-1}$ and $E_{\gamma,\rm iso}=1.65_{-1.16}^{+3.81} \times 10^{48}$ erg, respectively.

\begin{figure*}
	\centering
	\includegraphics[width=0.8\linewidth]{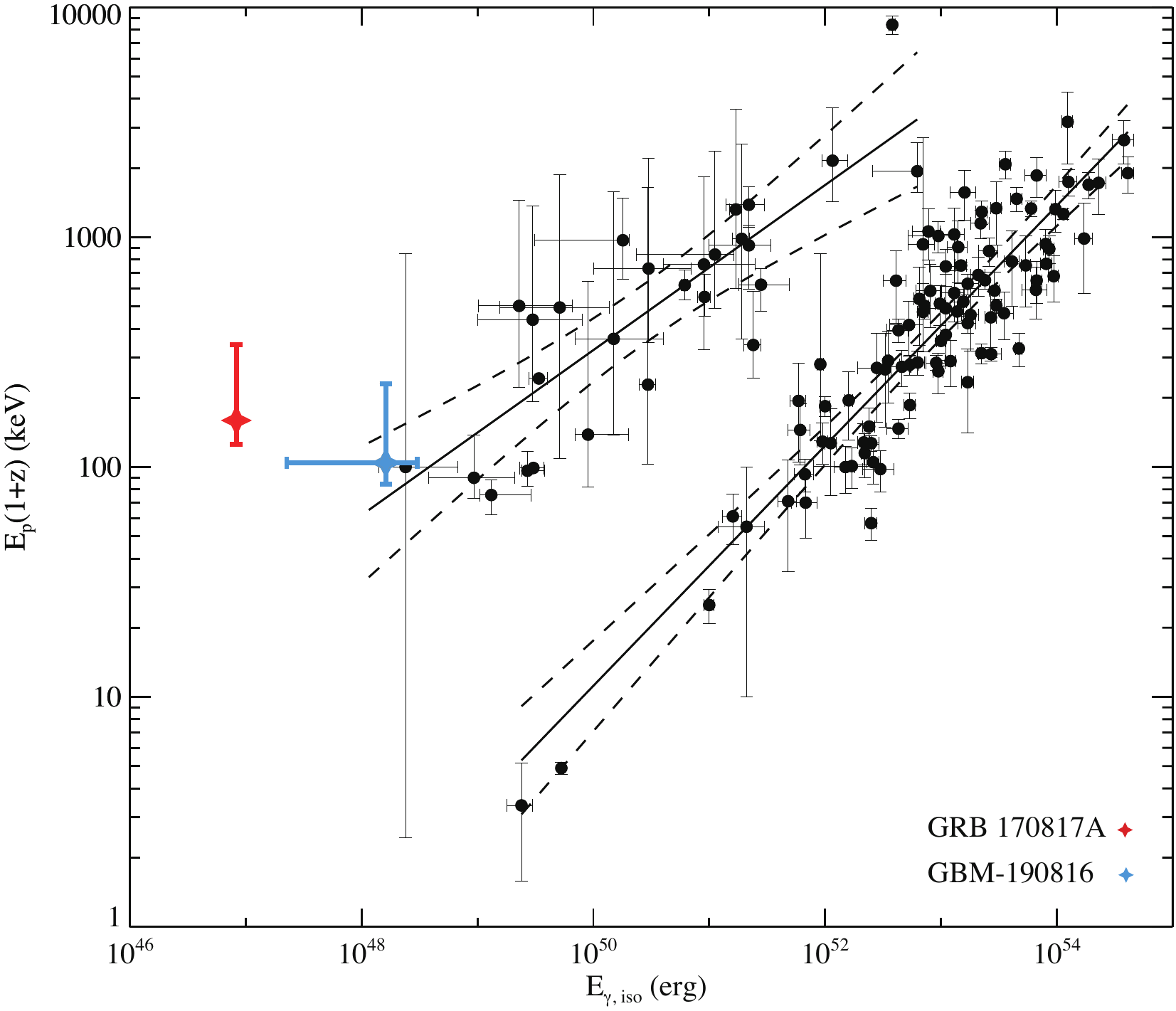}
	\caption{The $E_{\rm p}$ and $E_{\rm iso}$ correlation diagram. The red and blue stars represent GRB 170817A and GBM-190816, respectively. The upper and lower solid lines are the best-fit correlations for short and long GRB populations. All error bars in the panel denote the 1$\sigma$ uncertainties.}
	\label{fig:EpEiso}
\end{figure*}

(6) Amati relation. In order to check if GBM-190816 is an unusual event, we overplot GBM-190816 in the E$_p$-$E_{\rm \gamma,iso}$ correlation of all GRBs with known redshifts \citep{Amati2002,BZhang2009}. As shown in Figure \ref{fig:EpEiso}, unlike GRB 170817A, which is an outlier of the short GRB track, GBM-190816 is located well within the 1-$\sigma$ region of the short GRB population, suggesting that it is consistent with typical short GRBs in terms its spectral peak and total energy. 

A summary of the observed properties of GBM-190816 is listed in Table 2. The observed facts point towards the possibility that GBM-190816 is a short GRB with a sharp peak and typical temporal and spectral properties. The unusually short duration leads to a low fluence, which causes it being a sub-threshold event below the {\it Fermi}/GBM triggering threshold. 

\section{The possible sub-threshold Gravitational Wave Signal}

The sub-threshold GW signal associated with GBM-190816 was first announced through GCN Circular \citep{Goldstein2019}. The LIGO/Virgo Collaboration (LVC) did not announce this GW event on GraceDB\footnote{\url{http://gracedb.ligo.org/latest/}} as a significant candidate. As of writing, the GW data of GBM-190816 are not yet publicly available on Gravitational Wave Open Science Center\footnote{\url{https://www.gw-openscience.org/}}. However we can still obtain the following information about this event through the GCN Circular \citep{Goldstein2019} and the Gravitational-Wave Observatory Status website\footnote{\url{http://www.gw-openscience.org/summary\_pages/detector\_status/day/20190816/}}: 
\begin{enumerate}
 \item LIGO Hanford Observatory (H1) was not collecting data at that time so only Livingston Observatory (L1) and Virgo Observatory (V1) data are available. In any case, this event is a network detection rather than a single-interferometer detection.
 \item By applying the offline analysis of the data from L1 and V1, LVC identified a possible compact binary merger candidate at 2019-08-16 21:22:13.027 UTC (GPS time: 1250025751.027).
 \item 
 As a sub-threshold network detection event \citep{Goldstein2019}, the network S/N of this event is below the threshold of GW analysis pipelines, which is 12. According to the public O3 event GW190425's paper \citep{Abbott2020}, only events with the S/N higher than 4 will further calculate the FAR. So the network S/N of GBM-190816 should be between 4 and 12.
 \item The source localization was obtained by combining the L1-V1 data and the GRB data. The 90\% error of the source area corresponds to 5855 sq. deg. while the 50\% error of the source area is 1257 sq. deg. According the updated GCN Circular by the LIGO/Virgo/Fermi collaborations
 \citep{Shawhan2019} and the LALInference \citep{Veitch2015}, the 90\% and 50\% errors of the source area are down to 3219 sq. deg. 
 and 744 sq. deg., respectively. 
 \item The luminosity distance of the event is constrained to $428_{-143}^{+143}$ Mpc at 90\% CI \citep{Shawhan2019}.
 \item If the signal is astrophysical, the lighter compact object of this CBC event may have a mass $< 3 M_\odot$ \citep{Goldstein2019}.
\end{enumerate}

\begin{figure*}
	\centering
	\includegraphics[width=0.7\linewidth]{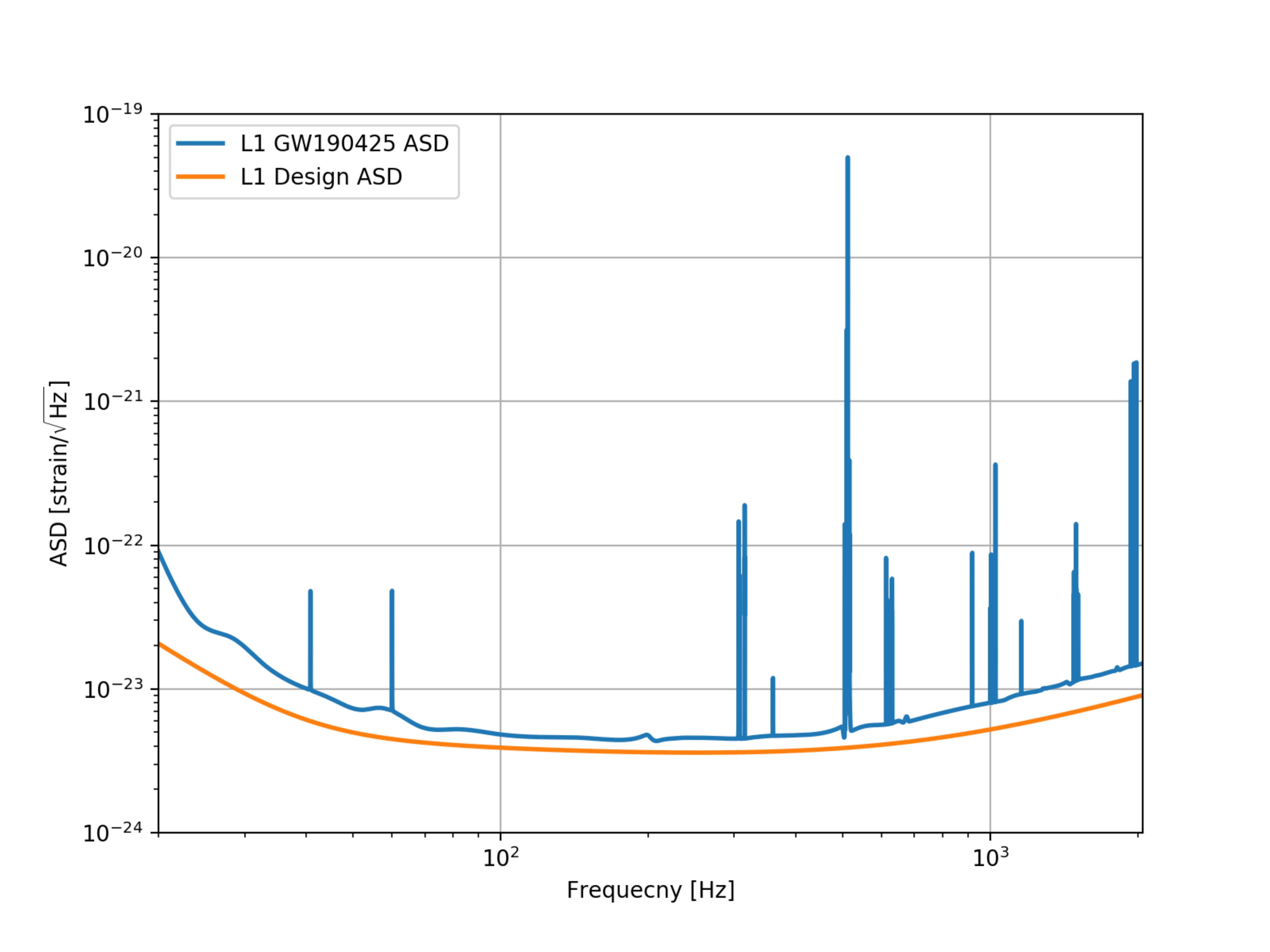}
	\caption{The GW190425 L1 ASD and the aLIGO L1 design ASD.}
	\label{fig:GBM-190816_mimic_ASD}
\end{figure*}

In order to constrain the mass ratio of the two objects in this GW event, the following three assumptions are made for simplicity: (1) One compact object of this CBC event is an NS with a mass of $1.4 M_\odot$. This is based on the information that the lighter compact object may have a mass $<$ 3 solar masses \citep{Goldstein2019} and that there is an associated putative GRB. (2) The L1 detector's sensitivity of GBM-190816 is the same as GW190425. Since GBM-190816's GW data are not public, we cannot use the actual data to calculate the Amplitude Spectral Density (ASD) of the detectors. On the other hand, the GW data of GW190425 are public now. Both GW190425 and GBM-190816 are quasi-single-detector events (both only have the L1 and V1 data, but the sensitivity of V1 is much worse than L1), the status of the detectors are public on the Gravitational-Wave Observatory Status website, which shows that their the sensitivities of L1 are almost the same\footnote{\url{https://www.gw-openscience.org/detector_status/day/20190425/}}\footnote{\url{https://www.gw-openscience.org/detector_status/day/20190816/}}. We use the official ASD of GW190425\footnote{\url{https://dcc.ligo.org/LIGO-P2000026/public}} to mimic the L1 sensitivity of GBM-190816, as shown in Figure \ref{fig:GBM-190816_mimic_ASD}. (3) The S/N of the event is 8 and mostly contributed by L1. This assumption is based on the fact that the NS-NS's inspiral range (smaller than horizon distance; to be discussed below) of V1 is much worse than L1. LVC's constraint on the luminosity distance is $428_{-143}^{+143}$ Mpc, which is much larger than V1's NS-NS detection range, so we assume that the S/N contributed by V1 is very small and the network S/N is almost contributed by L1. LVC defines a sub-threshold GW event with the network S/N below 12 and above 4 for network detections. We thus assume the S/N contributed by L1 is 8, which is the median value between 4 and 12, and is also the threshold S/N of a single detector for a confident GW candidate in network detections. Notice that for single-interferometer detections, the threshold S/N is larger than 8 \citep{Callister2017}. For real GW detections in O1/O2/O3, LVC set a threshold on FAR and $P_{astro}$, not directly on S/N. This can allow detection of events below the threshold S/N used in our paper. For a theoretical analysis, setting a threshold on S/N is a reasonable approach \citep{Abbott2019,Nitz2020}.

In the following, we demonstrate that by calculating the horizon distance of the L1 detector for different CBC GW signals with various mass ratios, we can constrain the mass ratio of GBM-190816 event to a specific range under the aforementioned assumptions. We assume that the orbital eccentricity at the merger is $\epsilon=0$ in following treatment. This is justified in view of the long-term decrease of $\epsilon$ due to gravitational wave radiation during the inspiral phase \citep[e.g.][]{Belczynski2002}. The method and equations follow the FINDCHIRP pipeline paper \citep{Allen2012}.

For a single GW detector, the location and orientation of the source are not easily obtained. Assume that the true distance of the GW source is $D$. It is more convenient to define an effective distance which combines the location and orientation of the source and is measurable, i.e. 
\begin{equation}
D_{\rm eff}= D\left[F_{+}^2 \left(\frac{1+\rm cos^{\rm 2}\rm \iota}{2}\right)^{2}+F_{\times}^2\rm cos^{\rm 2}\iota\right]^{-1/2},
\end{equation}
where $F_{+}$ and $F_{\times}$ are the detector's antenna responses to
the two polarization modes of the gravitational waveform and $\iota$ is the orientation of the GW source.

In the stationary phase approximation \citep{Sathyaprakash1991,Cutler1994,Poisson1995}, for $f$ $>$ 0, the frequency-domain GW waveform {\em in the inspiral stage} is
\begin{equation}
\tilde{h}(f)= -\left(\frac{5\pi}{24}\right)^{1/2}\left(\frac{G\mathcal{M}}{c^3}\right)\left(\frac{G\mathcal{M}}{c^2D_{\rm eff}}\right)\left(\frac{G\mathcal{M}}{c^3}\pi f\right)^{-7/6}e^{-i\Psi\left(f;M,\mu\right)}=\left(\frac{1Mpc}{D_{\rm eff}}\right)\mathcal{A}_{1 \rm Mpc}(\mathcal{M})f^{-7/6}e^{-i\Psi\left(f;M,\mu\right)} 
\end{equation}
where $\mathcal{M}$ is chirp mass and $M$ is total mass of the binary system,

\begin{equation}
\mathcal{A}_{1 \mathrm{Mpc}}(\mathcal{M})=-\left(\frac{5}{24 \pi}\right)^{1 / 2}\left(\frac{G M_{\odot} / c^{2}}{1 \mathrm{Mpc}}\right)\left(\frac{\pi G M_{\odot}}{c^{3}}\right)^{-1 / 6}\left(\frac{\mathcal{M}}{M_{\odot}}\right)^{5 / 6},
\end{equation}

\begin{equation}
 \Psi(f ; M, \mu)= 2 \pi f t_{0}-2 \phi_{0}-\pi / 4+\frac{3}{128 \eta}\left[v^{-5}+\left(\frac{3715}{756}+\frac{55}{9} \eta\right) v^{-3}\right.\left.-16 \pi v^{-2}+\left(\frac{15293365}{508032}+\frac{27145}{504} \eta+\frac{3085}{72} \eta^{2}\right) v^{-1}\right], 
\end{equation}
\begin{equation}
v=\left(\frac{G M}{c^{3}} \pi f\right)^{1 / 3},
\end{equation}
where the symmetric mass ratio
\begin{equation}
\eta=\frac{m_{1} m_{2}}{\left(m_{1}+m_{2}\right)^{2}}=\frac{\mu}{M}=\frac{q}{(1+q)^{2}},
\end{equation}
$q$ is the mass-ratio, and $\mu$ is the reduced mass
\begin{equation}
\mu=\frac{m_{1} m_{2}}{m_{1}+m_{2}}.
\end{equation}
We note that the effectively aligned spin parameter is very small for the detected merger events \citep{Abbott2019}, so it is ignored in our calculations.

For simplicity, we use the optimal S/N
\begin{equation}
 \tilde{\rho} = \sqrt{4 \int_{0}^{\infty} \frac{|\tilde{h}(f)|^{2}}{S_{n}(f)} d f} =\left(\frac{1 \rm Mpc}{D_{\rm eff}}\right) \sqrt{4 \mathcal{A}_{1 \rm Mpc}^{2}(\mathcal{M}) \int_{0}^{\infty} \frac{f^{-7 / 3}}{S_{n}(f)}df}
\end{equation}
to define the threshold S/N.
When the optimal S/N equals to the S/N threshold 8 (for a single GW detector in a network detection), we can calculate the horizon distance of a typical GW source, i.e. the farthest detection distance of a particular type of GW sources. For example, for CBCs (BH-BH, NS-NS or NS-BH mergers) we get
\begin{equation}
D_{\mathrm{hor}}=\frac{1 \rm{Mpc}}{\tilde{\rho}} \sqrt{4 \mathcal{A}_{1 \rm{Mpc}}^{2}(\mathcal{M}) \int_{0}^{\infty} \frac{f^{-7 / 3}}{S_{n}(f)} d f}.
\end{equation}

If we fix the mass of one compact object and change the mass ratio $q$, we can get different horizon distances as a function of $q$. Here we fix one object's mass to $1.4 M_\odot$ (the NS) and use the mimicked ASD as mentioned before. The lower frequency limit 20 Hz and the upper frequency limit 
\begin{equation}
f_{\text {isco }}=\frac{c^{3}}{6 \sqrt{6} \pi G M}
\end{equation}
are adopted in the integration.

\begin{figure*}
	\centering
	\includegraphics[width=0.7\linewidth]{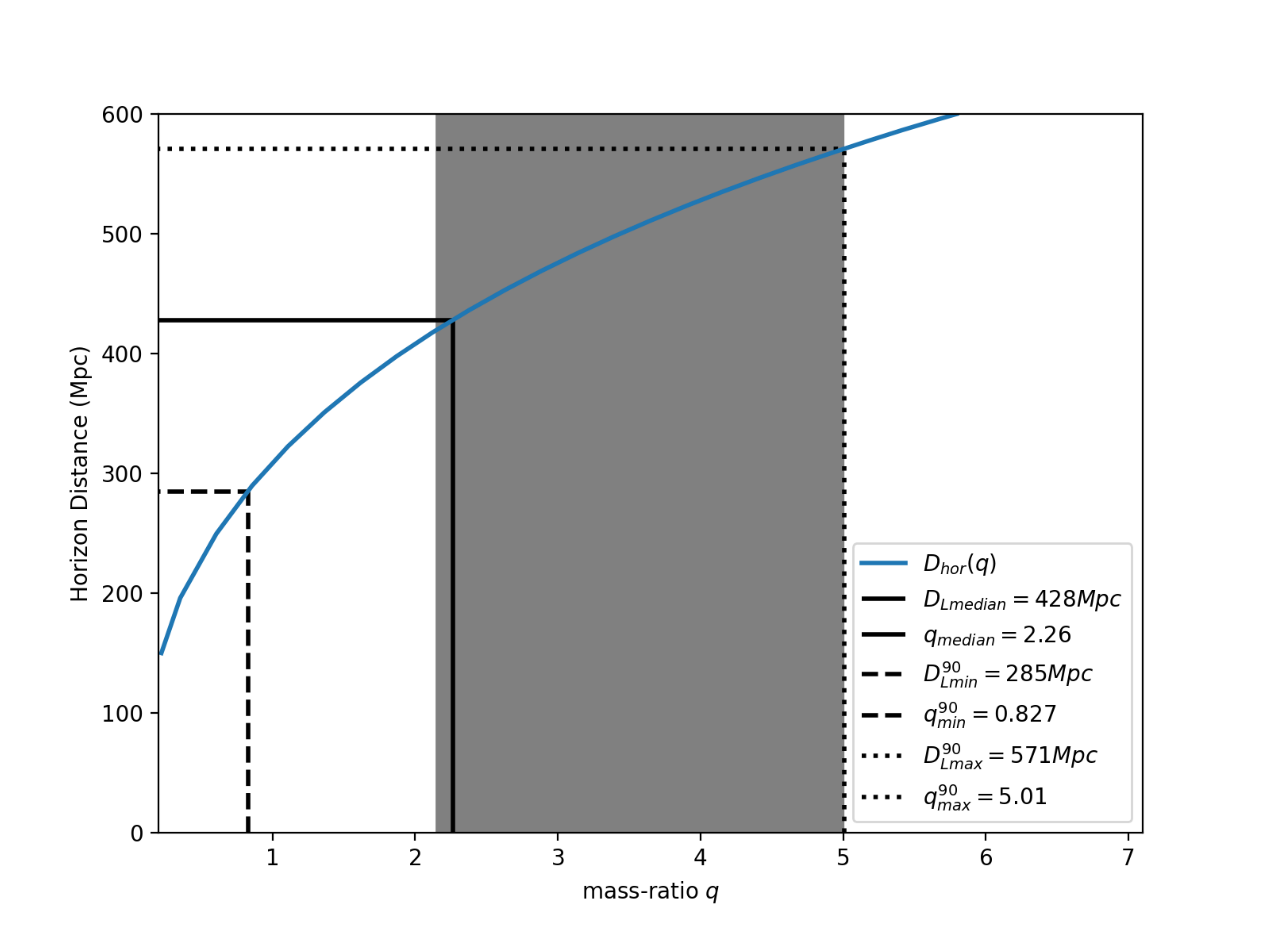}
	\caption{The inferred mass ratio $q$ range based on the available GW information (chosen optimal SNR is 8). The dashed and dotted lines represent the lower and upper limits (90\% CI) for $q=2.26_{-1.43}^{+2.75}$ constrained with the reported luminosity distance range. The gray area indicates that $q=2.26_{-0.12}^{+2.75}$ can be derived if the heavier compact object has a mass $>$ 3 solar masses.}
	\label{fig:distance_q}
\end{figure*}

The results are shown in Figure \ref{fig:distance_q}. Since the luminosity distance of this sub-threshold event is $428_{-143}^{+143}$ Mpc, we can utilize the upper and lower limits (90\% CI) of the luminosity distance to get the upper and lower limits (90\% CI) of the mass ratio. Based on the significant digits of the luminosity distance given by the GCN Circular, we keep 3 significant digits in the mass ratio $q$. We can constrain the mass ratio $q$ to $q=2.26_{-1.43}^{+2.75}$ with 90\% CI. Considering the fact that only the lighter compact object has a mass $< 3 M_\odot$, which indicates that the mass ratio $q$ should be $>$ 3/1.4 under the aforementioned assumption, we can further derive the mass ratio $q$ to $q=2.26_{-0.12}^{+2.75}$. This is displayed in the gray area in the Figure \ref{fig:distance_q}.

\begin{figure*}
	\centering
	\includegraphics[width=0.7\linewidth]{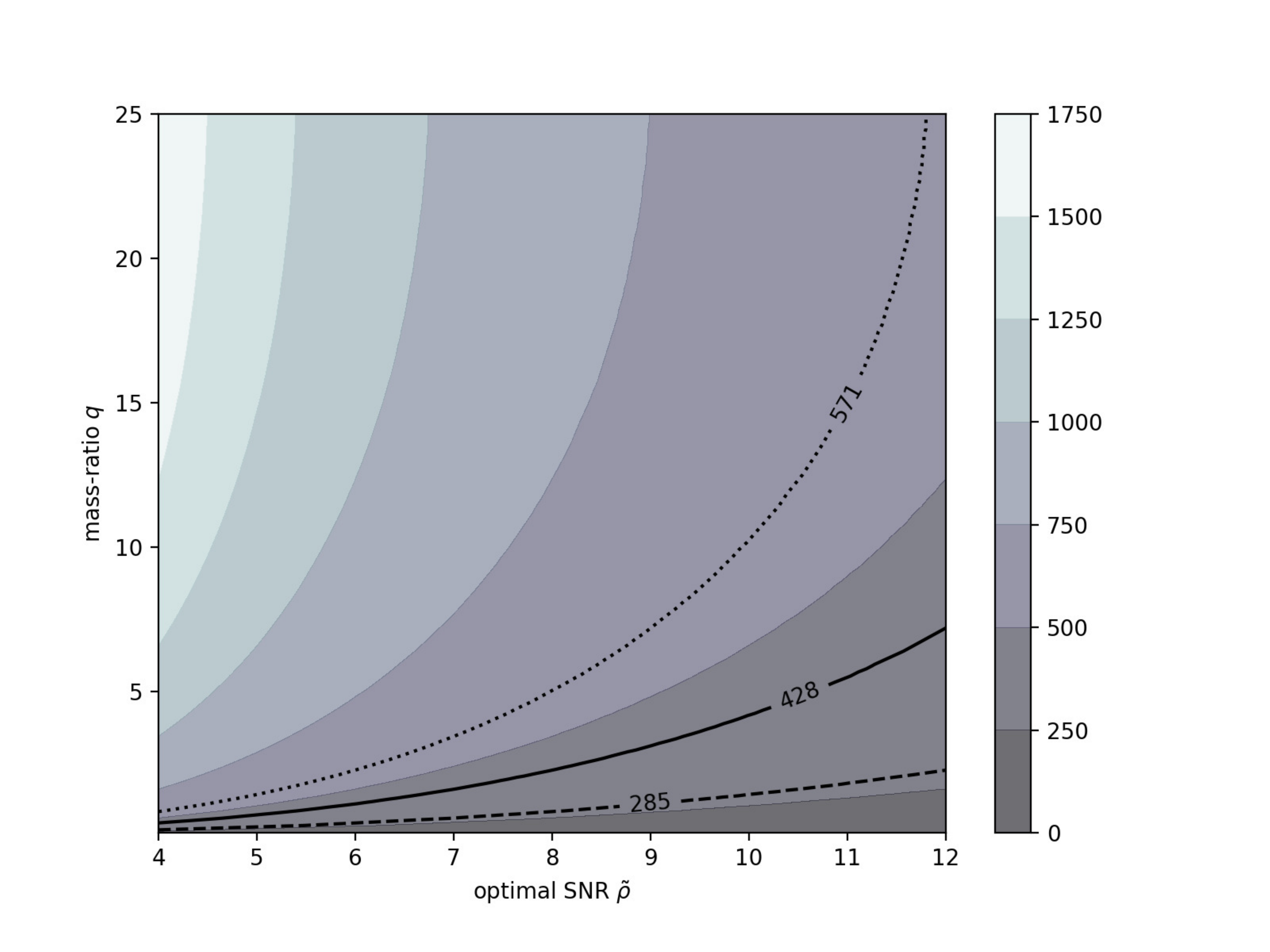}
	\caption{The inferred mass ratio $q$ range on different chosen optimal SNR. The
	dashed and dotted lines represent the lower and upper limits (90\% CI) of the reported luminosity distance. The solid line represents the median value of the reported luminosity distance. Gray area represents contours of the horizon distance at given SNR and $q$.}
	\label{fig:horizon_contor}
\end{figure*}

We also calculate the limit of the mass ratio when the S/N takes different thresholds, as shown in Figure \ref{fig:horizon_contor}. The color represents the value of the horizon distance at a given S/N and mass ratio. The solid, dashed and dotted lines represent the median, lower and upper limits (90\% CI) of the public luminosity distance, respectively. When the S/N is 8, the result returns to Figure \ref{fig:distance_q}. It is worth noting that when the S/N threshold becomes larger, the interval of the mass ratio becomes larger, which is contrary to the experience of standard gravitational wave Bayesian parameter estimation. The reason is that here we fix the interval width of the luminosity distance, which should become narrower when S/N becomes higher. So we take the median value of the S/N range to avoid this effect.

\section{Physical implications of the Electromagnetic Signal GBM-190816}

In this section, assuming that both the sub-threshold GW and the sub-threshold GRB are real and are also related, we discuss the physical implications of such an association. 

The leading model of short GRBs invokes a black hole central engine surrounded by a hyper-accreting torus. For GBM-190816 and its putative GW counterpart, the most likely possibility is an NS-BH merger with a not-too-large mass ratio $q$, so that the NS is tidally disrupted before the merger and there is neutron-rich material outside the BH event horizon after the merger to power the short GRB. We discuss this possibility in \S4.1. 
The allowed wide range of $q$ from the available GW data actually allows a ``plunging'' NS-BH merger (i.e. the NS is not tidally disrupted but is swallowed as a whole by the BH) \citep{Shibata2009} or even a BH-BH merger (the maximum NS mass is likely smaller than $3 M_\odot$. In both of these scenarios, a short GRB with a short delay with respect to the GW may demand more exotic scenarios such that at least one member of the merger system is charged \citep{BZhang2016,BZhang2019a,Dai2019}. We discuss this possibility and constrain the model parameters in \S4.2. Finally, in \S4.3, we generally discuss the physical implication of the 1.57 s delay between the putative GW event and the putative short GRB event.

\subsection{NS-BH Merger with Tidal Disruption: Constraints on Model Parameters}\label{sec:NS-BH}

For NS-BH mergers, whether or not there is matter left outside the post-merger BH event horizon is determined by a comparison between the tidal disruption radius $d_{\rm tidal}$ and the radius of the innermost stable circular orbit $R_{\rm ISCO}$ \citep{Shibata2009}. In general, the total mass $M_{\rm out}$ of the matter left outside the BH event horizon after tidal disruption of the NS can be divided into two components: the disc mass $M_{\rm disc}$ and the dynamical ejecta mass $M_{\rm dyn}$.
Numerical simulations suggest that $M_{\rm out}$ depends on the mass ($M_{\rm BH}$) and the dimensionless spin ($\chi_{\rm BH}$) of the BH, the baryonic mass of the NS ($M_{\rm NS}^{\rm b}$), as well as the tidal deformability ($\Lambda_{\rm NS}$) of the NS, i.e. \citep{fou18}
\begin{equation}
M_{\mathrm{out}}=M_{\mathrm{NS}}^{\mathrm{b}}\left[\max \left(\alpha \frac{1-2 \rho}{\eta^{1 / 3}}-\beta \tilde{R}_{\mathrm{ISCO}} \frac{\rho}{\eta}+\gamma, 0\right)\right]^{\delta},
\label{eq:M_out}
\end{equation}
where $\eta=q /(1+q)^{2}$, $\rho=\left(15 \Lambda_{\mathrm{NS,1.4}}\right)^{-1 / 5}$ ($\Lambda_{\mathrm{NS,1.4}}$ represents $\Lambda_{\mathrm{NS}}$ for the NS mass at $1.4M_{\odot}$), and the dimensionless ISCO radius follows
\begin{equation}
 \tilde{R}_{\mathrm{ISCO}}=R_{\mathrm{ISCO}} c^{2} / G M_{\mathrm{BH}}=3+Z_{2}-\operatorname{sgn}\left(\chi_{\mathrm{BH}}\right) \sqrt{\left(3-Z_{1}\right)\left(3+Z_{1}+2 Z_{2}\right)},
\end{equation}
 with $Z_{1}=1+\left(1-\chi_{\mathrm{BH}}^{2}\right)^{1 / 3}\left[\left(1+\chi_{\mathrm{BH}}\right)^{1 / 3}+\left(1-\chi_{\mathrm{BH}}\right)^{1 / 3}\right]$
and $Z_{2}=\sqrt{3 \chi_{\mathrm{BH}}^{2}+Z_{1}^{2}}$ \citep{bard72}. The empirical parameters yield to $\alpha=0.308$, $\beta=0.124$, $\gamma=0.283$, and $\delta=1.536$. 

The dynamical ejecta mass $M_{\rm dyn}$ depends on $M_{\mathrm{BH}}$, the NS gravitational mass $M_{\mathrm{NS}}$, $M_{\mathrm{NS}}^{\mathrm{b}}$, $\mathcal{\chi}_{\mathrm{BH}}$, the NS compactness $C_{\mathrm{NS}}=\sum_{k=0}^{2} a_{k}^{c}\left(\ln \Lambda_{\mathrm{NS,1.4}}\right)^{k}$ \citep[``C-Love" relation,]{yagi17}, and the angle between the BH spin and the binary total angular momentum $\iota_{\mathrm{tilt}}$:
\begin{equation}
 M_{\mathrm{dyn}}=M_{\mathrm{NS}}^{\mathrm{b}}\{ \max \left[a_{1} q^{n_{1}}\left(1-2 C_{\mathrm{NS}}\right) / C_{\mathrm{NS}}
-a_{2} q^{n_{2}} \tilde{R}_{\mathrm{ISCO}}\left(\chi_{\mathrm{eff}}\right)
\right.\left.\left.+a_{3}\left(1-M_{\mathrm{NS}} / M_{\mathrm{NS}}^{\mathrm{b}}\right)+a_{4}, 0\right]\right\},
\label{eq:M_dyn}
\end{equation}
\normalsize where $\chi_{\mathrm{eff}}=\chi_{\mathrm{BH}} {\rm cos}\iota_{\mathrm{tilt}}$ is the effective BH spin, with empirical parameters being $a_{1}=4.464 \times 10^{-2}$, $a_{2}=2.269 \times 10^{-3}$, $a_{3}=2.431$, $a_{4}=-0.4159$, $n_{1}=0.2497$,
and $n_{2}=1.352$, respectively \citep{kaw16}. For simplicity, we adopt cos$\iota_{\mathrm{tilt}}$=1 in this paper. The NS baryonic mass $M_{\rm NS}^{\rm b}$ is related to its gravitational mass $M_{\rm NS}$ by
$M_{\mathrm{NS}}^{\mathrm{b}}=M_{\mathrm{NS}}\left(1+\frac{0.6 C_{\mathrm{NS}}}{1-0.5 C_{\mathrm{NS}}}\right)$ \citep{latt01}\citep[see also][]{gao19}. As pointed out in \cite{barb19}, the maximal dynamical ejecta mass $M_{\rm dyn,max}$ cannot exceed $0.5M_{\rm out}$. We thus assume $M_{\rm dyn,max}=0.3M_{\rm out}$ which is consistent with the result from numerical simulations of NS-BH mergers in the near-equal-mass regime \citep{fou19}.
The disc mass $M_{\rm disc}$ is obtained by combing Eqs. (\ref{eq:M_out}) and (\ref{eq:M_dyn}):
\begin{equation}
M_{\mathrm{disc}}= M_{\mathrm{out}}-M_{\mathrm{dyn}}.
\label{eq:M_disc}
\end{equation}

We consider a relativistic jet launched from the central engine through the BZ
mechanism. The kinetic energy of the jet may be calculated by\footnote{The detailed derivation of this equation invokes the approximated BZ luminosity \citep{tch10} and a rough estimation of the duration of disk accretion \citep{barb19}. More generally, the relation between the kinetic energy and the disk mass should be non-linear. The linear relation in Equation (\ref{eq:E_k,jet}) could be a first order approximation. }
\begin{equation}
E_{\mathrm{K}, \mathrm{jet}}=\epsilon\left(1-\xi_{\mathrm{w}}\right) M_{\mathrm{disc}} c^{2} \Omega_{\mathrm{H}}^{2} f\left(\Omega_{\mathrm{H}}\right),
\label{eq:E_k,jet}
\end{equation}
where $\epsilon$ is a dimensionless constant that depends on the ratio of the magnetic energy density to disc pressure at saturation \citep{haw15},
$\xi_{\mathrm{w}}$ is the fraction of energy that goes to the disk wind (rather than the jet) which is related to the kilonova power, $\Omega_{\mathrm{H}}=\frac{\chi_{\mathrm{BH,f}}}{2(1+\sqrt{1-\chi_{\mathrm{BH,f}}^{2}})}$ is the dimensionless angular velocity evaluated at the BH horizon,
and $f\left(\Omega_{\mathrm{H}}\right)=1+1.38 \Omega_{\mathrm{H}}^{2}-9.2 \Omega_{\mathrm{H}}^{4}$ is a correction factor for high-spin
values. 

The dimensionless spin of the final BH remnant, $\chi_{\mathrm{BH,f}}$, is related to the initial BH spin $\chi_{\rm BH}$ in the NS-BH binary through \citep{buo08,pann13} 
\begin{equation}
\chi_{\mathrm{BH,f}}=\frac{\chi_{\mathrm{BH}} M_{\mathrm{BH}}^{2}+l_{z}\left(\bar{r}_{\mathrm{ISCO} }, \chi_{\mathrm{BH,f}}\right) M_{\mathrm{BH}} M_{\mathrm{NS}}}{M^{2}},
\label{eq:x_BH,f}
\end{equation}
where $M=M_{\rm BH}+M_{\rm NS}$, and 

\begin{equation}
l_{z}\left(\bar{r}_{\mathrm{ISCO}},\chi_{\mathrm{BH,f}}\right)={\rm sgn}(\chi_{\rm BH,f}) \frac{\bar{r}_{\rm ISCO}^{2} -2{\rm sgn}(\chi_{\rm BH,f}) \chi_{\rm BH,f} \sqrt{\bar{r}_{\rm ISCO}}+\chi_{\rm BH,f}^{2}}{\sqrt{\bar{r}_{\rm ISCO}}\left(\bar{r}_{\rm ISCO}^{2}-3 \bar{r}_{\rm ISCO} +2{\rm sgn}(\chi_{\rm BH,f})\chi_{\rm BH,f} \sqrt{\bar{r}_{\rm ISCO}}\right)^{1 / 2}}
\end{equation}
is the orbital angular momentum per unit mass of a test particle orbiting the BH remnant at the
ISCO, $\bar{r}_{\rm ISCO}$ is similar to $\tilde{R}_{\mathrm{ISCO}}$ but replaces $\chi_{\rm BH}$ by $\chi_{\rm BH,f}$. Equation (\ref{eq:x_BH,f}) in the geometric units is same as that in the normalized units. For simplicity, the rotation of the NS, the mass and angular momentum of the tidal material, as well as the GW radiation was not taken into account in equation (\ref{eq:x_BH,f}).

In order to connect the BH accretion power with the observed GRB power, we assume a Gaussian-shape structured jet \citep{Beniamini2019} with an angular distribution of the kinetic energy and Lorentz factor $\Gamma$ following
\begin{equation}
\frac{d E}{d \Omega}(\theta) =E_{\mathrm{c}} e^{-\left(\theta / \theta_{\mathrm{c,j}}\right)^{2}}, ~~
\Gamma(\theta) =\left(\Gamma_{\mathrm{c}}-1\right) e^{-\left(\theta / \theta_{\mathrm{c,j}}\right)^{2}}+1,
\label{eq:Ec}
\end{equation}
where $E_{\mathrm{c}}=E_{\mathrm{K}, \mathrm{jet}} / \pi \theta_{\mathrm{c,j}}^{2}$. Such a structure was long proposed as the typical GRB structured jet \citep{Zhang2002} and has been successfully applied to model GW170817/GRB 170817A \citep{Lazzati2018,Lyman2018,Troja2019,ghir19,Ryan2019}.

At the viewing angle $\theta_v$, the isotropic gamma-ray radiation energy can be estimated as
\begin{equation}
E_{\gamma,\mathrm{iso}}\left(\theta_{v}\right) \simeq \eta_{\gamma} \int \frac{D_{\rm p}^{3}}{\Gamma} \frac{d E}{d \Omega} d \Omega,
\label{eq:E_iso}
\end{equation}
where $\eta_{\gamma}$ the efficiency to convert the EM luminosity to the radiation luminosity in $\gamma$-ray band, $D_{\rm p}=1/[\Gamma(1-\beta {\rm cos}\alpha)]$ is the Doppler factor, and ${\rm cos}\alpha=\cos \theta_{{v}} \cos \theta+\sin \theta_{{v}} \sin \theta \cos \varphi$. 

\begin{figure*}
\includegraphics[angle=0,width=0.5\textwidth]{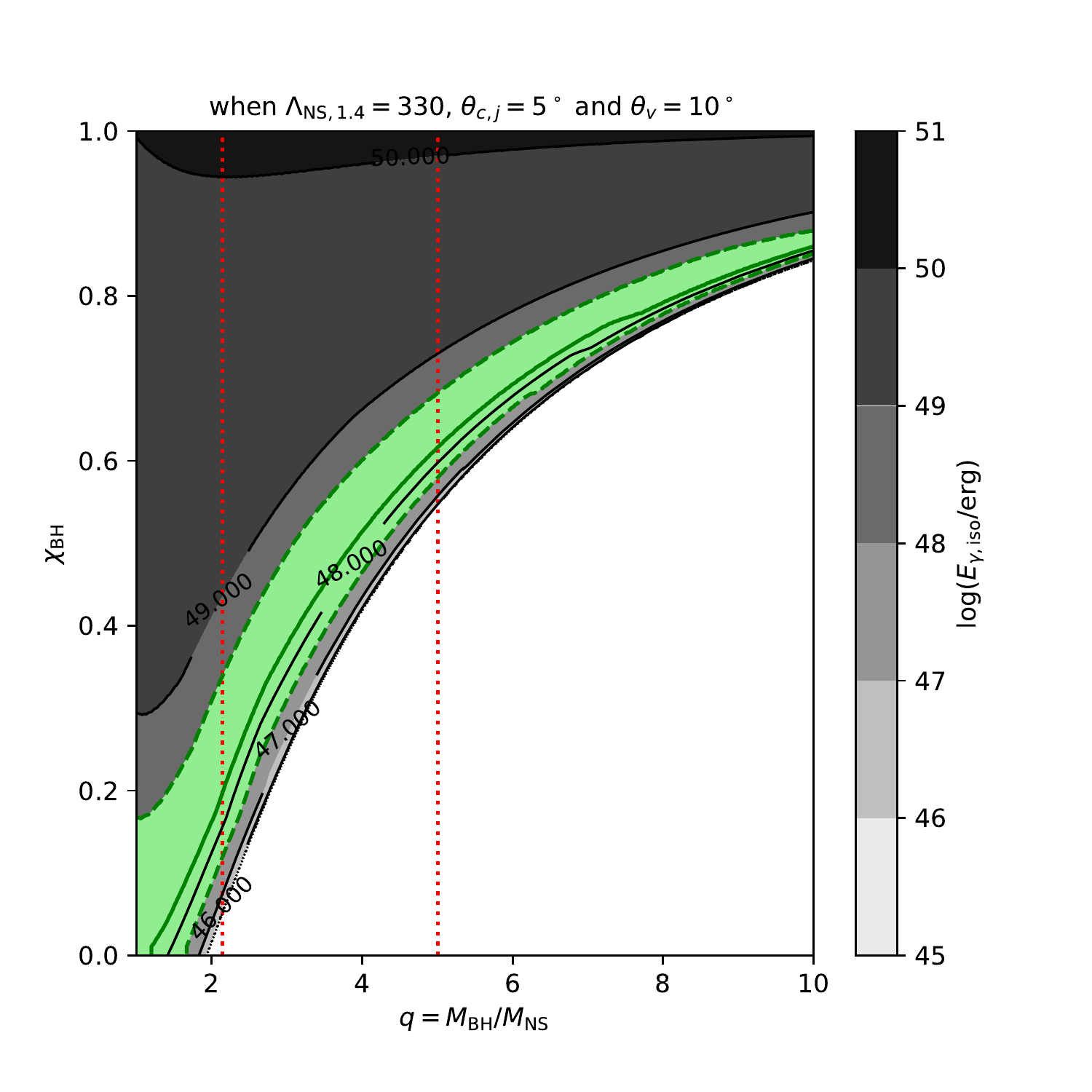}
\includegraphics[angle=0,width=0.5\textwidth]{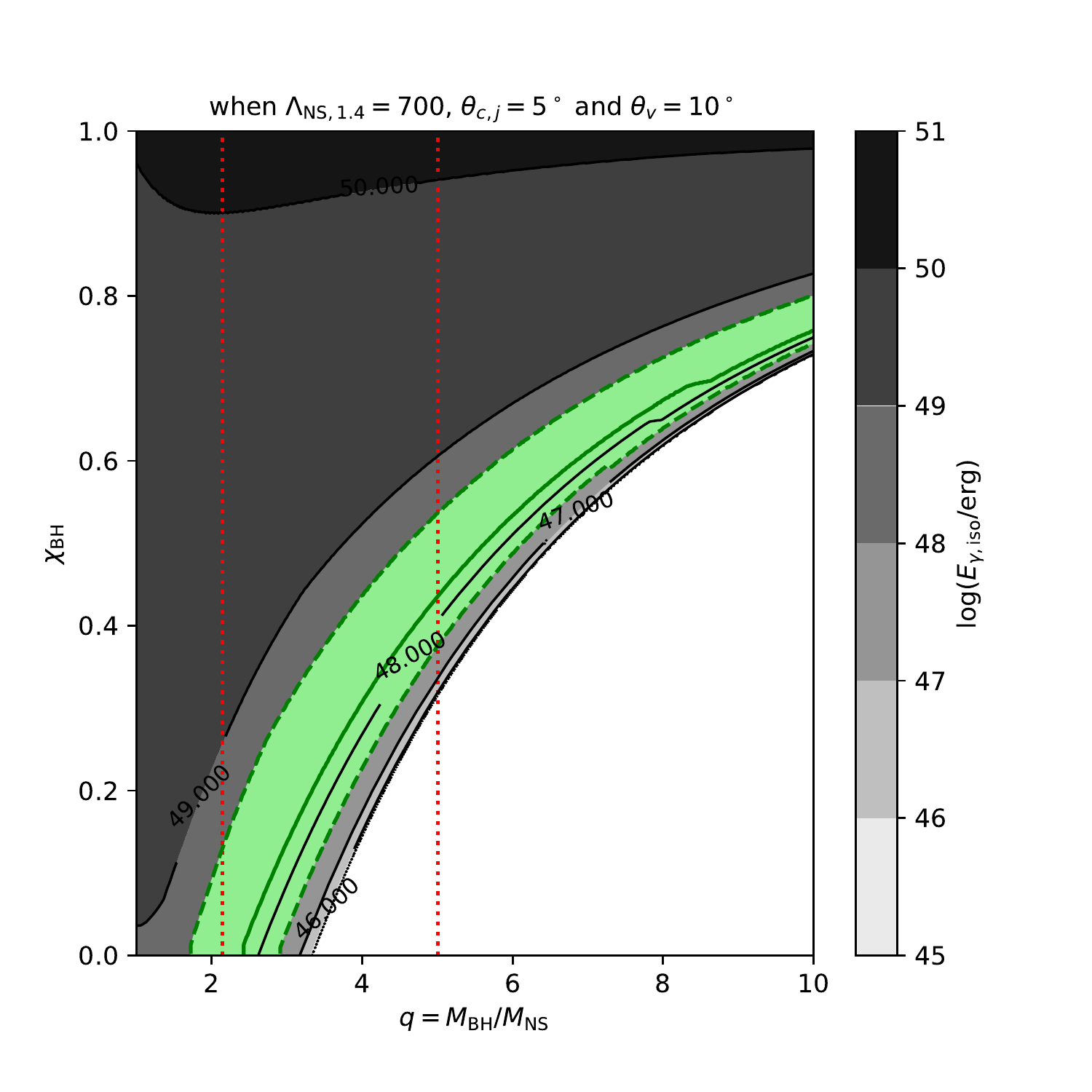}
\includegraphics[angle=0,width=0.5\textwidth]{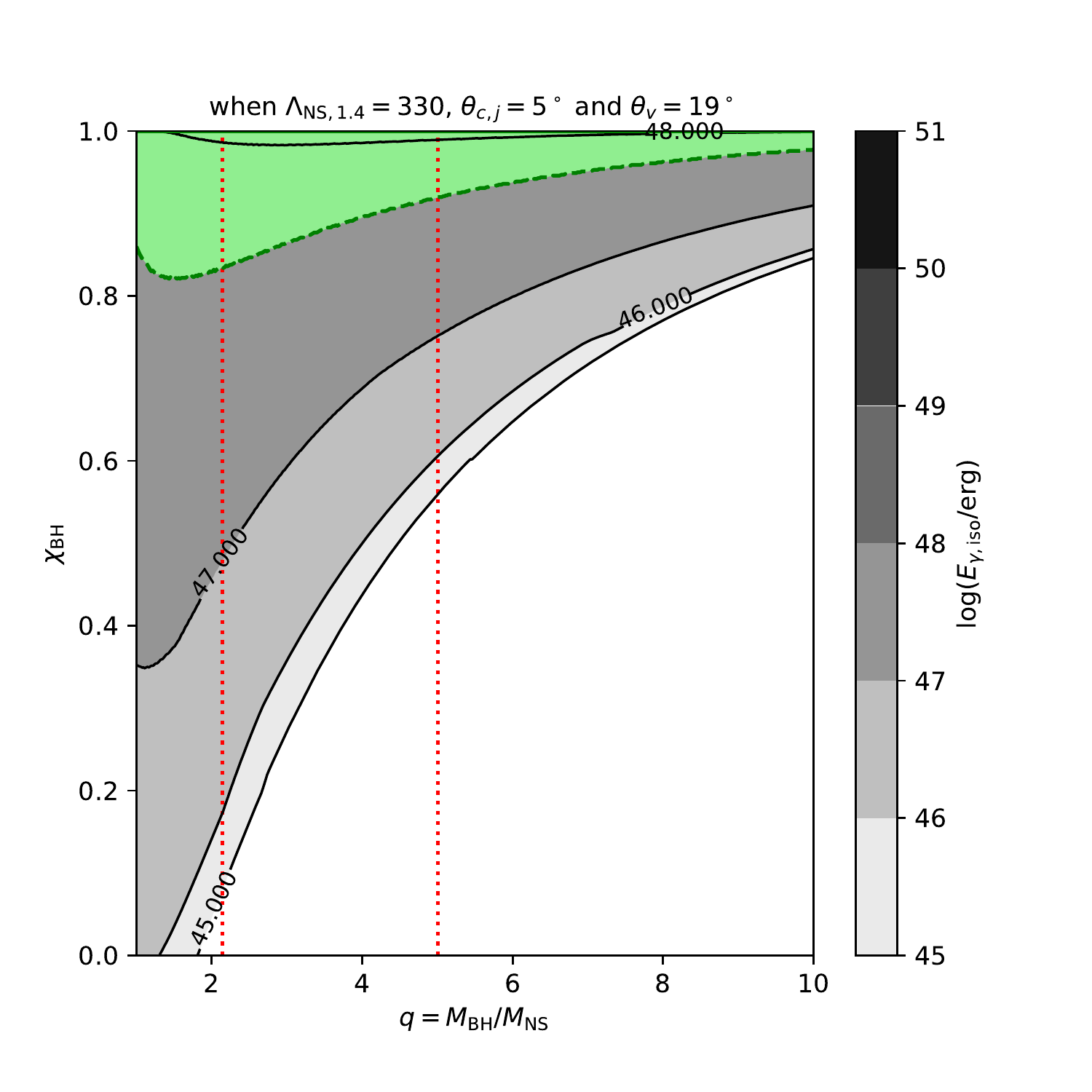}
\includegraphics[angle=0,width=0.5\textwidth]{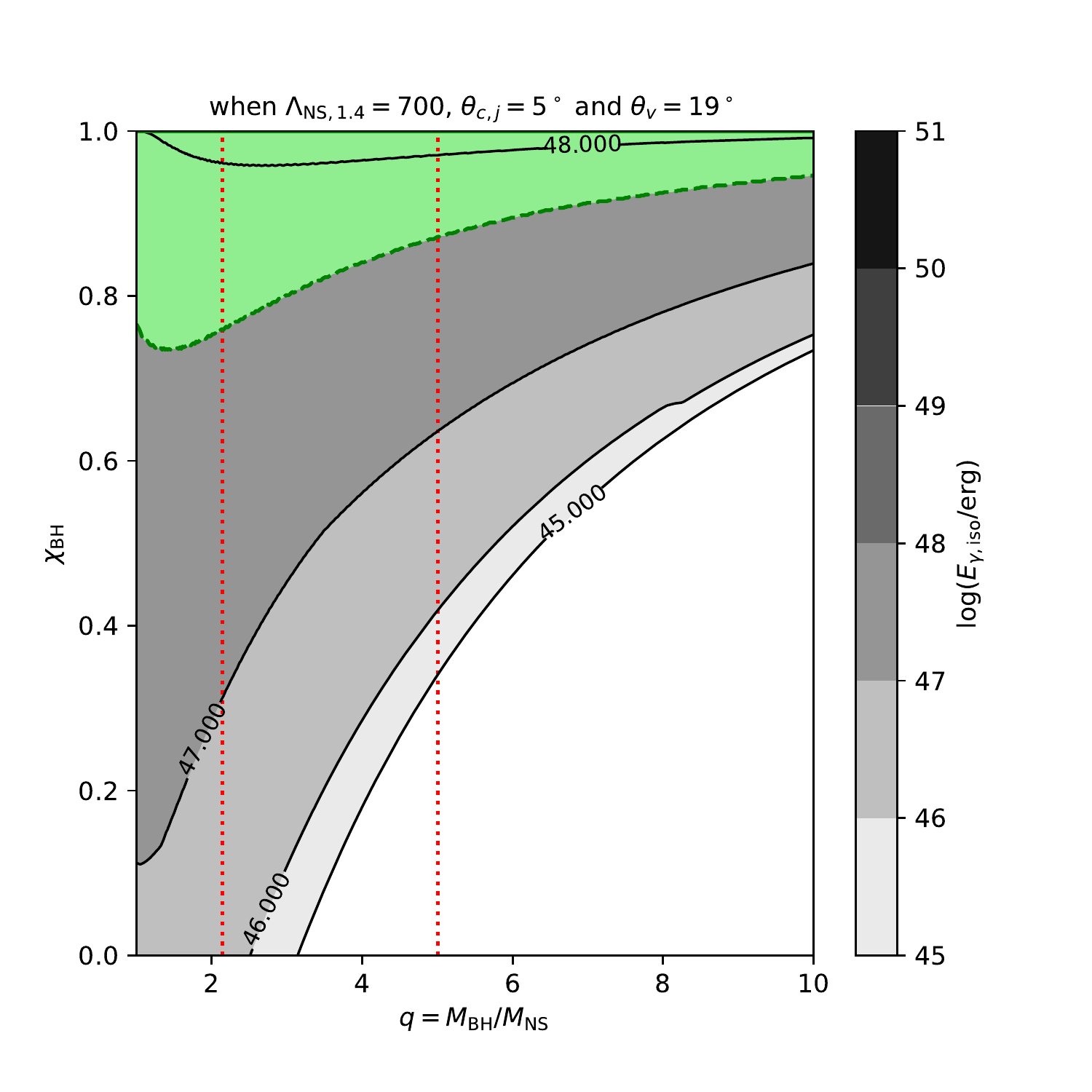}
\caption{The isotropic gamma-ray radiation energy $E_{\gamma,\rm iso}$ in the $q-\chi_{\rm BH}$ parameter space assuming a Gaussian-shaped jet with a narrow jet core $\theta_{\rm c,j}=5^\circ$ and various values of $\theta_{\rm v}$ and $\Lambda_{\rm NS}$ (as marked). Two NS EOS (SFHo and DD2) are assumed for $\Lambda_{\rm NS,1.4}=330$ and $700$. The filled green regions represent the allowed parameter space that can reproduce the E$_{\gamma,\rm iso}$ of GBM-190816. The red lines indicate the GW constraints on $q$.}
\label{fig:ejecta1}
\end{figure*}

\begin{figure*}
\includegraphics[angle=0,width=0.5\textwidth]{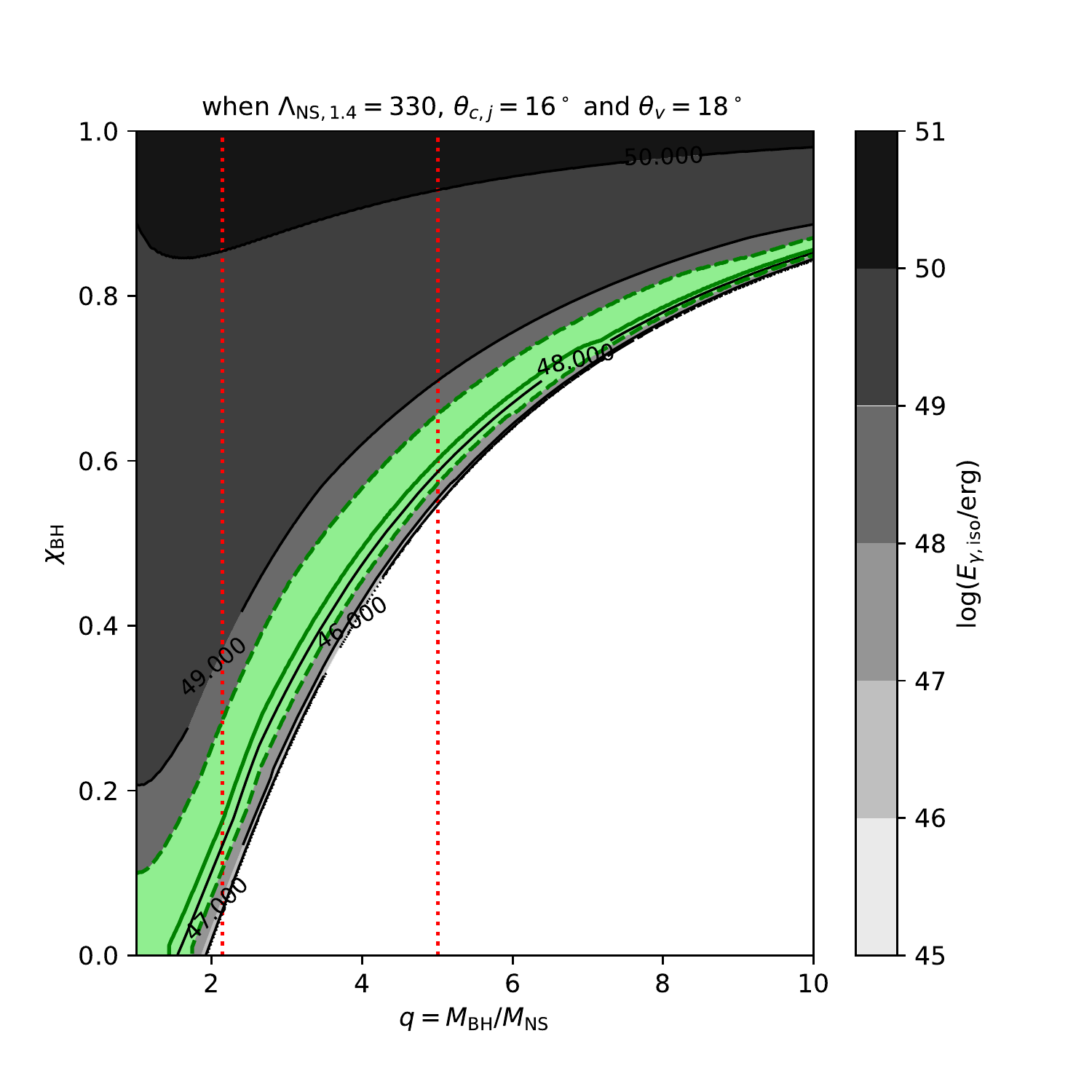}
\includegraphics[angle=0,width=0.5\textwidth]{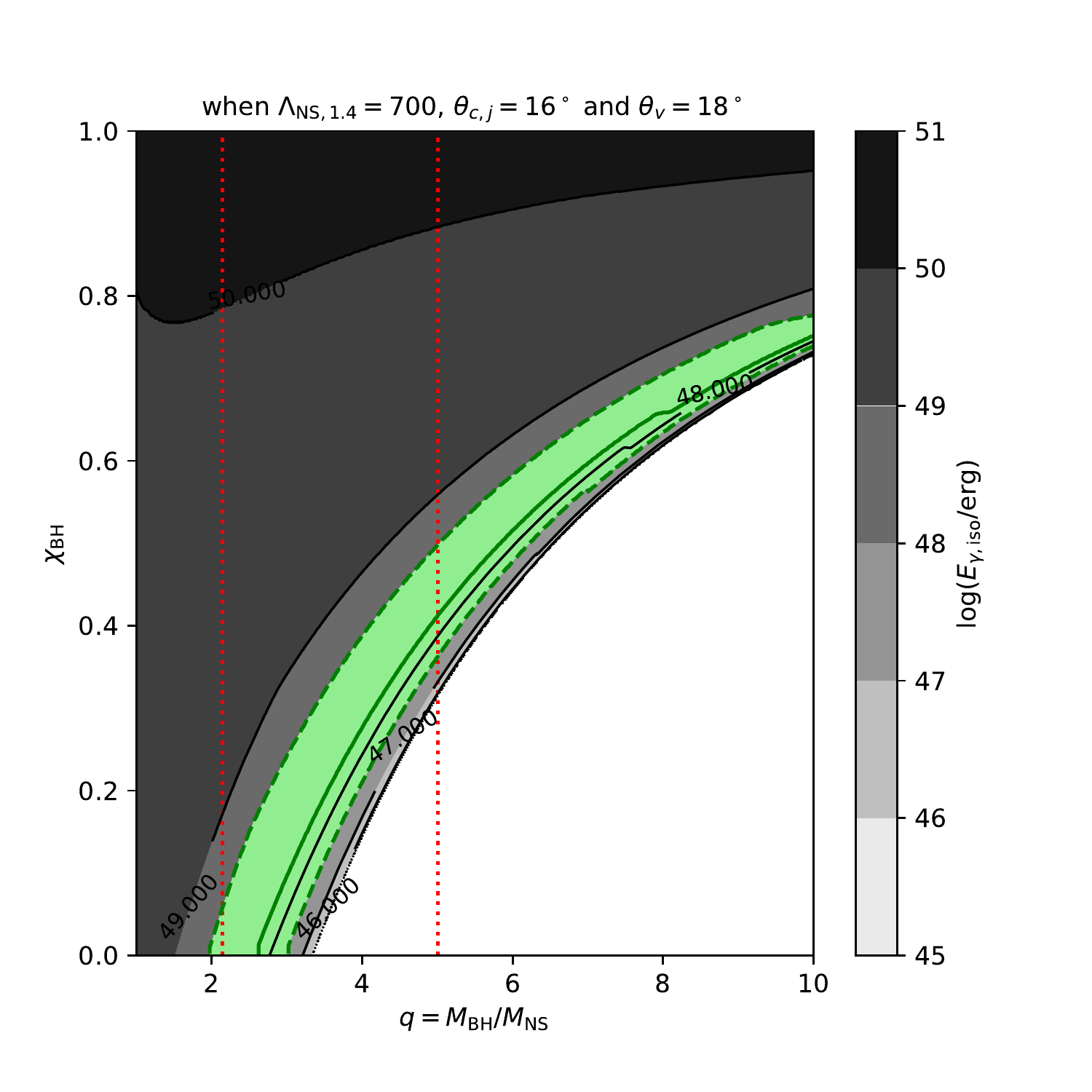}
\includegraphics[angle=0,width=0.5\textwidth]{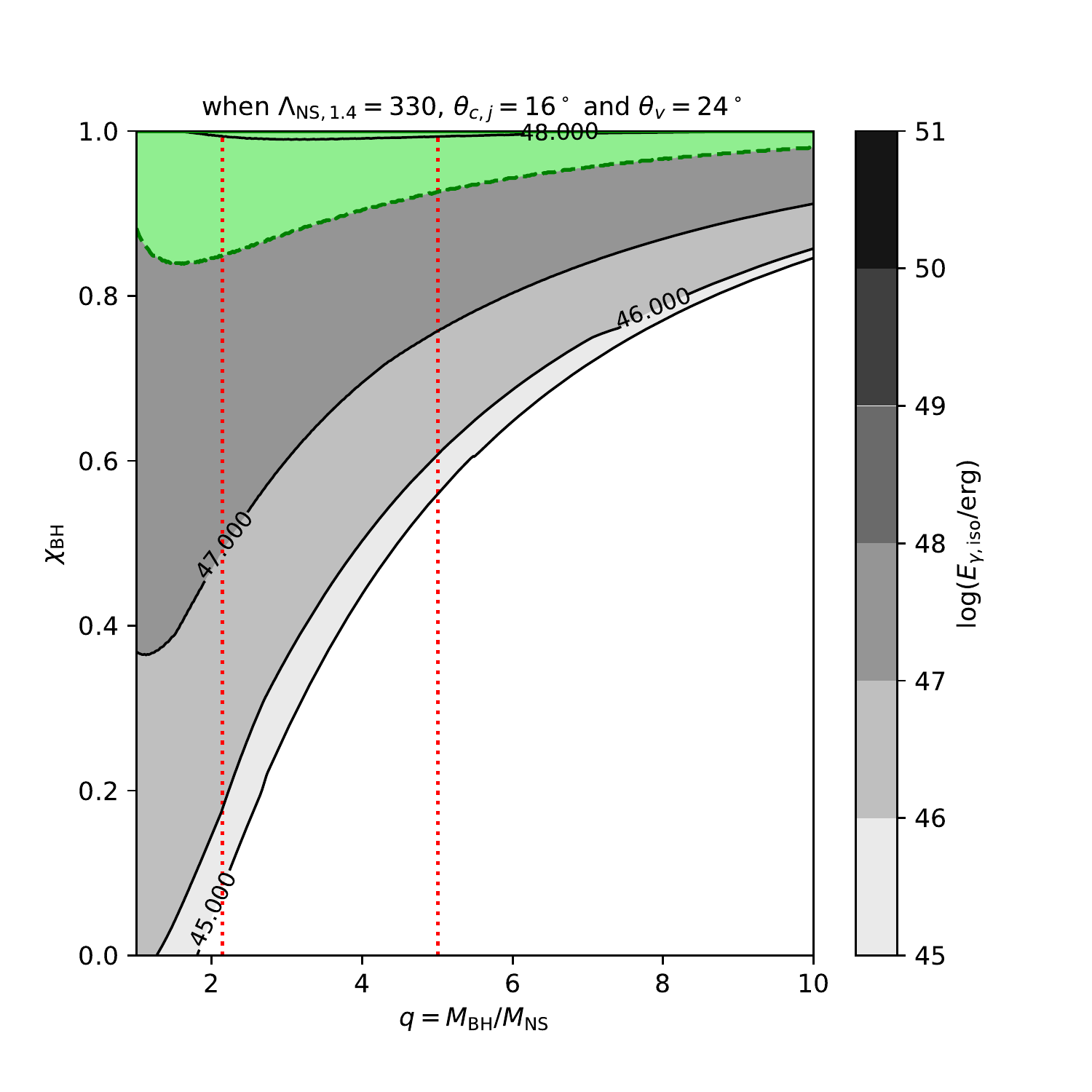}
\includegraphics[angle=0,width=0.5\textwidth]{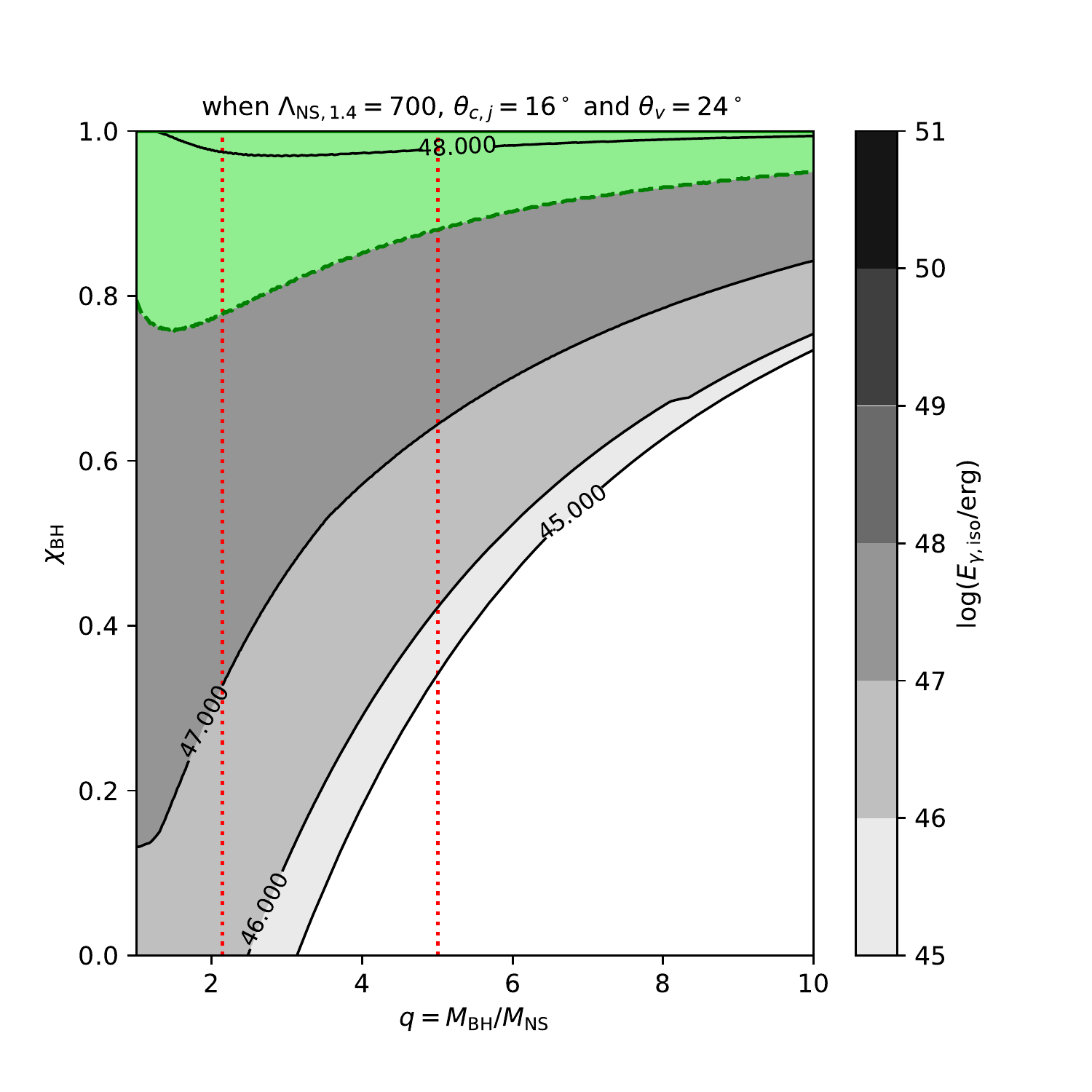}
\caption{Same as Figure \ref{fig:ejecta1} but for the case of a wide jet with core $\theta_{\rm c,j}=16^\circ$ and various values of $\theta_{\rm v}$, $\Lambda_{\rm NS}$ (as marked). } 
\label{fig:ejecta2}
\end{figure*}

Combining equations (\ref{eq:M_out})-(\ref{eq:E_iso}), we can calculate the isotropic radiation energy of the jet $E_{\gamma,\rm iso}$ as a function of some parameters (e.g., $q$ and $\chi_{\rm BH}$) of the BH and the NS under certain assumptions. Owing to the limited information of this event, we have to assume the event possess some typical characteristics of short GRBs, e.g. $\epsilon=0.015$, $\xi_{\mathrm{w}}=0.01$, $\eta_{\gamma}=10\%$, $M_{\rm NS}=1.4~M_{\odot}$, and $\Gamma_{\rm c}=100$ \citep[see][]{barb19}. Since we do not know the NS equation of state, we take $\Lambda_{\rm NS,1.4}=330$ (SFHo EoS) and 700 (DD2 EoS) to cover a range of possible cases. Furthermore, we consider two example cases for a narrow jet core with $\theta_{\rm c,j}=5^{\circ}$ and a wide jet core with $\theta_{\rm c,j}=16^{\circ}$. The former value is motivated by GW170817/GRB 170817A \citep{ghir19} while the latter is consistent with the claimed opening angle of some observed short GRBs \citep{fong15}. Our constraints on $q$ and $\chi_{\rm BH}$ for different cases are presented in Figures \ref{fig:ejecta1} and \ref{fig:ejecta2}. Despite the flexible allowed range of $q$ and $\chi_{\rm BH}$, our results suggest that the viewing angle should lie in the most possible range in order to achieve observed $E_{\gamma,\mathrm{iso}}$, which is $10^{\circ}-19^{\circ}$ ($18^{\circ}-24^{\circ}$) for the narrow (wide) jet core cases, respectively, as shown in Figure \ref{fig:ejecta1} (Figure \ref{fig:ejecta2}). In addition, different $\Lambda_{\rm NS,1.4}$ values (corresponding to different NS EoSs) can visibly influence the green regions in the $q$-$\chi_{\rm BH}$ plane achieving the observed $E_{\gamma,\mathrm{iso}}$, but it does not significantly change the most possible allowed ranges of the viewing angle.

The constraints presented in Figs. \ref{fig:ejecta1} and \ref{fig:ejecta2} are by no means definite. This is because from merger parameters to the observed GRB parameters, there are three major steps in modeling, which involve many unknown parameters: First, the uncertainties in the NS-BH merger physics may introduce a large error in the disc mass. Second, jet launching from a BH-disc system in principle involves two possible mechanisms: neutrino-anti-neutrino annihilation \citep{pop99} or Blandford-Znajek (BZ)
mechanism \citep{bland77} and we only considered the latter mechanism. Furthermore, we have adopted a simple analytical formula to denote the BZ power. Third, the radiation efficiency (which depends on the energy dissipation site and mechanism, e.g. photosphere emission, internal shocks, or magnetic dissipation) and geometry (jet structure and viewing angle) are essential to determine how a BZ-powered jet is observed as a short GRB. As explained above, when deriving $(q, \chi_{\rm BH})$ presented in Figs. \ref{fig:ejecta1} and \ref{fig:ejecta2}, we have adopted the typical values for some parameters based on the best known models. Allowing broader distributions of these parameters would further weaken the constraints.

\subsection{Plunging NS-BH Merger or BH-BH Merger: Constraints on Charge in the cCBC Systems }

For an NS-BH merger with a relatively large $q$ (e.g., $\sim$ 5), the NS would plunge into the BH as a whole.  Alternative mechanisms \citep[e.g.,][]{mcw11,tsang12,do16,levin18,BZhang2019a,Dai2019,pan19,zhong19} have to be introduced to explain the observed GRB. One group of the mechanisms which recently receive increasing interest are the electric and magnetic dipole radiation, magnetic reconnection, and BZ mechanism of the charged objects in the binary system. 

The charged compact binary coalescence (cCBC) models involve at least one member of the binary carries either a constant \citep{BZhang2016,BZhang2019a} or increasing \citep{levin18,Dai2019} charge. The high-energy EM emission can be produced either before \citep{BZhang2019a,Dai2019} or after \citep{pan19,zhong19} the merger.

\subsubsection{cCBC with a Constant Charge}
Here we consider the simplest case in which both objects carry a constant charge, which are denoted as $Q_1$ and $Q_2$, respectively. The following derivation applies to both plunging NS-BH and BH-BH scenarios. 

Two components can contribute to the EM luminosity of such a constant charge binary. The electric dipole radiation \citep{Deng2018} component reads \citep{BZhang2019a}
\begin{equation}
L_{\mathrm{e}, \mathrm{dip}}=\frac{1}{6} \frac{c^{5}}{G}\left(\hat{q}_{1}^{2}+\hat{q}_{2}^{2}\right)\left(\frac{r_{s}\left(m_{1}\right)}{a}\right)^{2}\left(\frac{r_{s}\left(m_{2}\right)}{a}\right)^{2},
\end{equation}
where $a$ is the semi-major axis with the eccentricity $e = 0$ assumed, $m_1$ and $m_2$ are the masses of compact objects, $\hat{q}_i \equiv Q_i/Q_{c,i}$ (i = 1, 2) are the dimensionless charges, $Q_{c,i}\equiv 2 \sqrt{G}m_i$ are the critical charges \citep{BZhang2016}, and $r_s(m_i)$ are the Schwarzschild radii of the two merging objects. 

Following \cite{BZhang2016,BZhang2019a}, the magnetic dipole radiation luminosity reads 
\begin{equation}
L_{\mathrm{B}, \mathrm{dip}} =\frac{196}{1875} \frac{c^{5}}{G}\left(\frac{\hat{q}_{1} m_{1}+\hat{q}_{2} m_{2}}{M}\right)^{2} \times\left(\frac{r_{s}\left(\mu\right)}{a}\right)^{4}\left(\frac{r_{s}(M)}{a}\right)^{11}. 
\end{equation}

Since at the final moment of the merger, the global open field lines in the binary system cover almost the full sky \citep{BZhang2016} and since there is no matter outside the BH event horizon to collimate the Poynting flux outflow, the estimated EM luminosity is the isotropic equivalent one:
\begin{equation}
L_{\gamma,\rm iso}=\eta_{\gamma}\left(L_{\mathrm{e}, \mathrm{dip}} +L_{\mathrm{B}, \mathrm{dip}}\right).
\end{equation}
For the most optimistic cases, we assumed $\eta_{\gamma} \sim 1$.

For an NS-BH merger system, at least the NS is charged \citep{Michel1982,BZhang2019a}. We adopt the following simplest assumptions: (1) only the NS carries a constant charge; (2) the NS mass is $1.4 M_{\odot}$; (3) $a=a_{min} = r_{s}(m_{BH}) + 2.4r_{s}(m_{NS})$ ($r_{NS}$ = 2.4 $r_s$ for neutron star) at the merger time; (4) mass ratio $q$ is $q=2.26_{-0.12}^{+2.75}$, which is constrained by the GW signal. We can then obtain that $\hat{q}_{NS}$ is $1.495_{-0.001}^{+0.210} \times 10^{-4} $. Consequently, the absolute charge $Q_{NS}$ is $2.162_{-0.002}^{+0.302} \times 10^{26} $ e.s.u. The dimensionless charge of a NS can be estimated as 
\citep{BZhang2019a}
\begin{equation}
\hat q_{\rm NS} \simeq \frac{3 \Omega B_p R^3}{2 c \sqrt{G} M} \cos\alpha = (4.4 \times 10^{-4
}) B_{15} P_{-3}^{-1} R_6^3 M_{1.4}^{-1} \cos\alpha.
\end{equation}
In order to satisfy the observational constraint, one requires ${B_{15}}/{P_{-3}}\sim 0.340_{-0.001}^{+0.047}$. This implies that the neutron star has to be a millisecond magnetar before the merger. The condition to form such a magnetar in BNS mergers is contrived, so this scenario is disfavored.

Similarly, for a charged BH-BH merger system we adopt the following two most straightforward assumptions: (1) the lighter BH has a mass of $2.8~M_{\odot}$, which is less than $3~M_{\odot}$ and falls into the BH mass regime; (2) only the lighter BH carries a constant dimensionless charge $\hat q$ (for the same absolute charge $Q$, a lighter BH carries a higher $\hat q$ which is more relevant). The mass ratio $q$ of this system should be different from the the range constrained above assuming an NS-BH merger, but this ratio does not enter the problem in view of assumption (2) above. We constrain the black hole charge as $\hat{q}_{BH} = 7.308_{-0.147}^{+3.909} \times 10^{-5}$ and the corresponding absolute charge $Q_{BH} = 2.114_{-0.042}^{+1.131} \times 10^{26} $ e.s.u. The demanded dimensionless charge is comparable to the one required to explain the putative $\gamma$-ray event\citep{Connaughton16} associated with the the first BH-BH merger event \citep{BZhang2016}. Contrived conditions are again needed for a BH to carry such a large charge.

\subsubsection{cCBC with an Increasing Charge}

\begin{figure*}
\includegraphics[angle=0,width=0.5\textwidth]{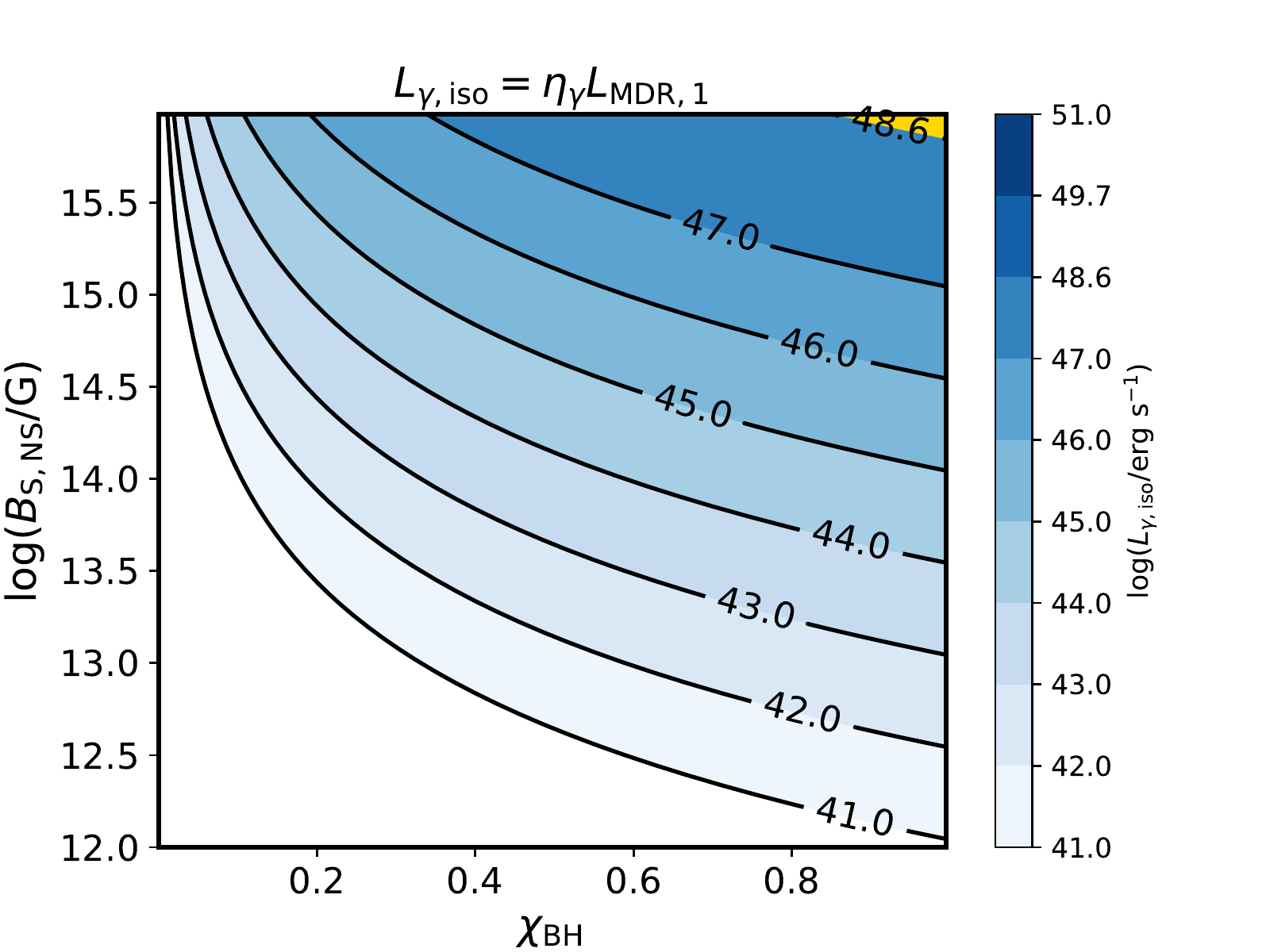}
\includegraphics[angle=0,width=0.5\textwidth]{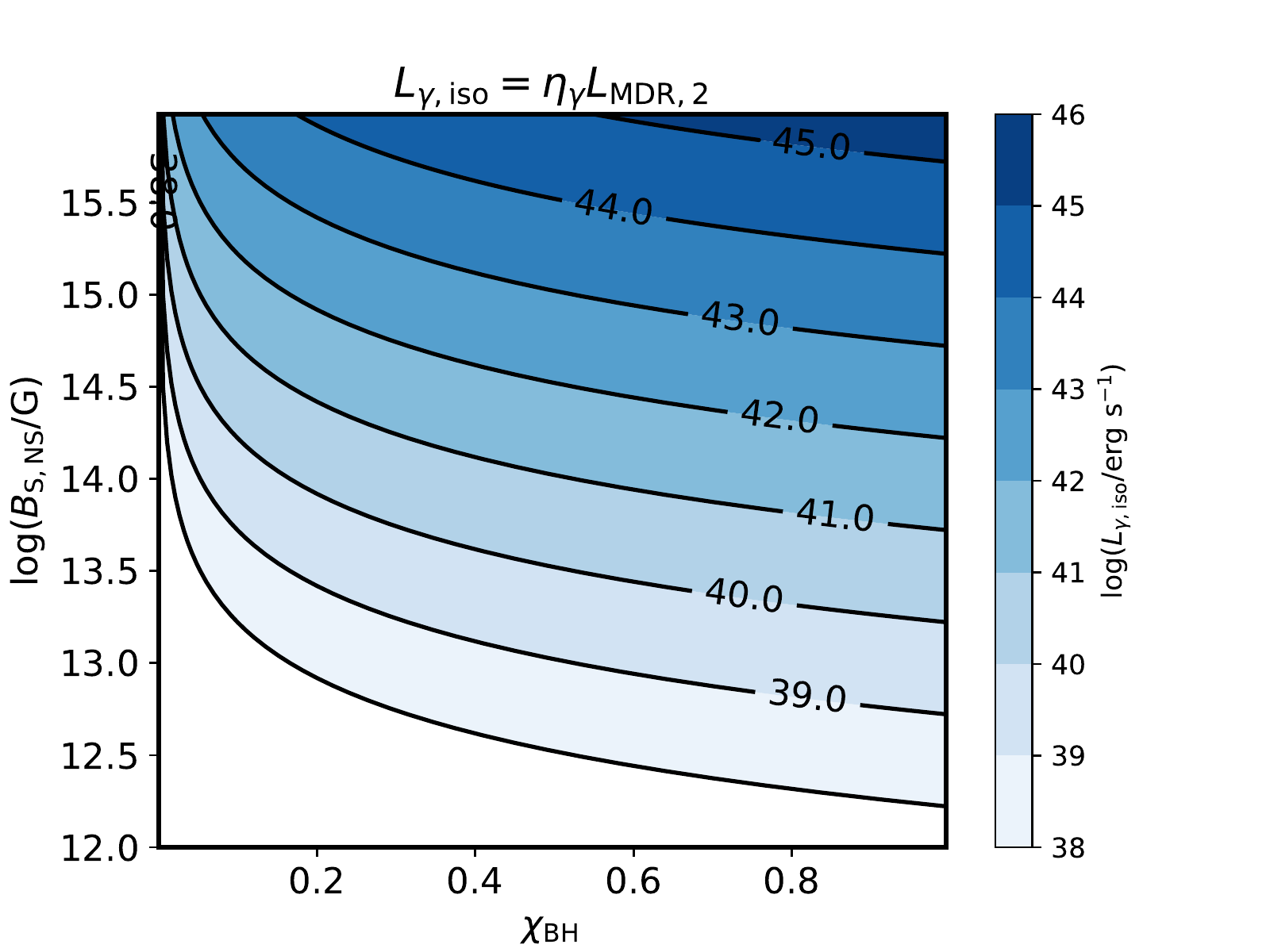}
\includegraphics[angle=0,width=0.5\textwidth]{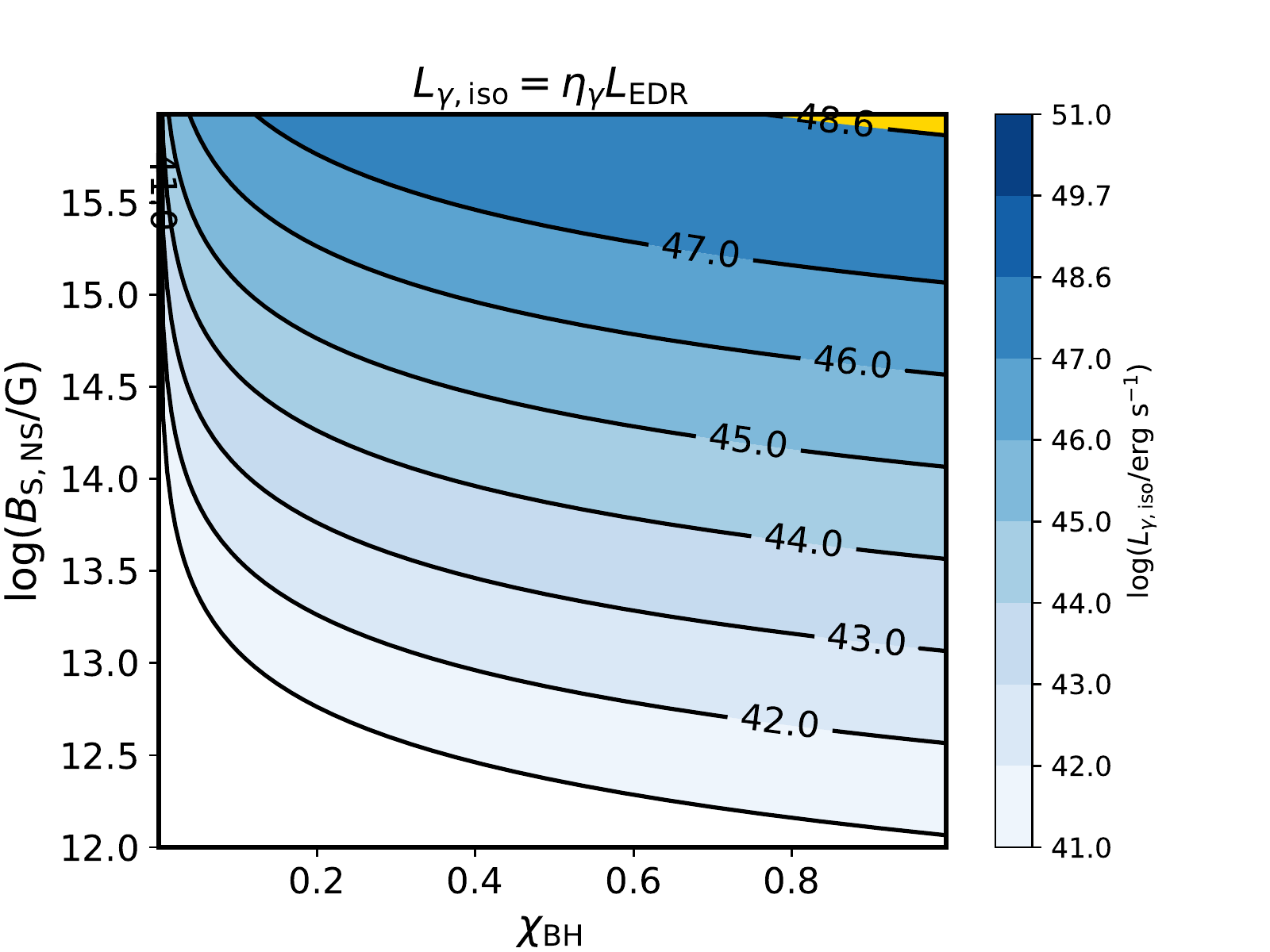}
\includegraphics[angle=0,width=0.5\textwidth]{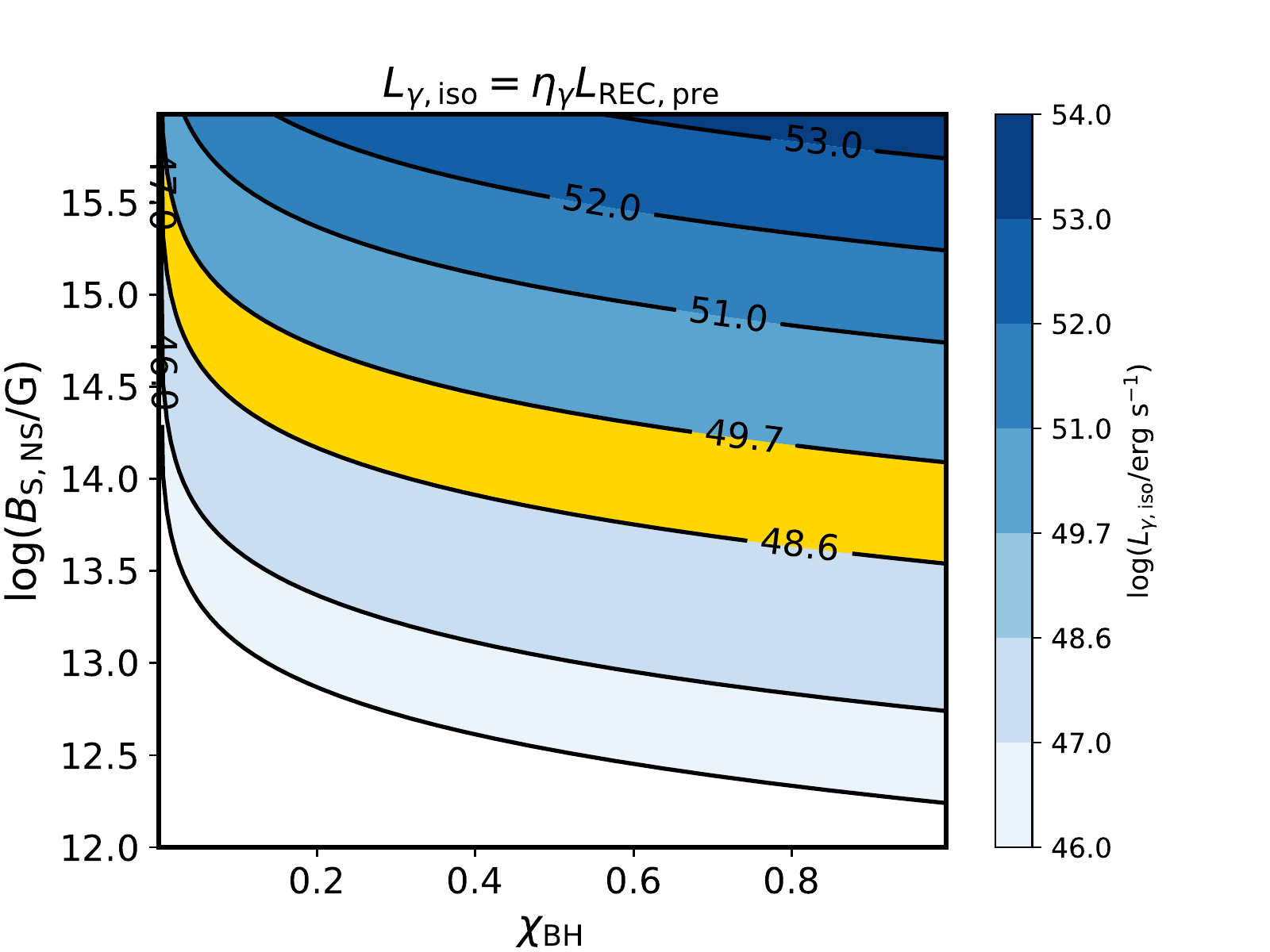}
\includegraphics[angle=0,width=0.5\textwidth]{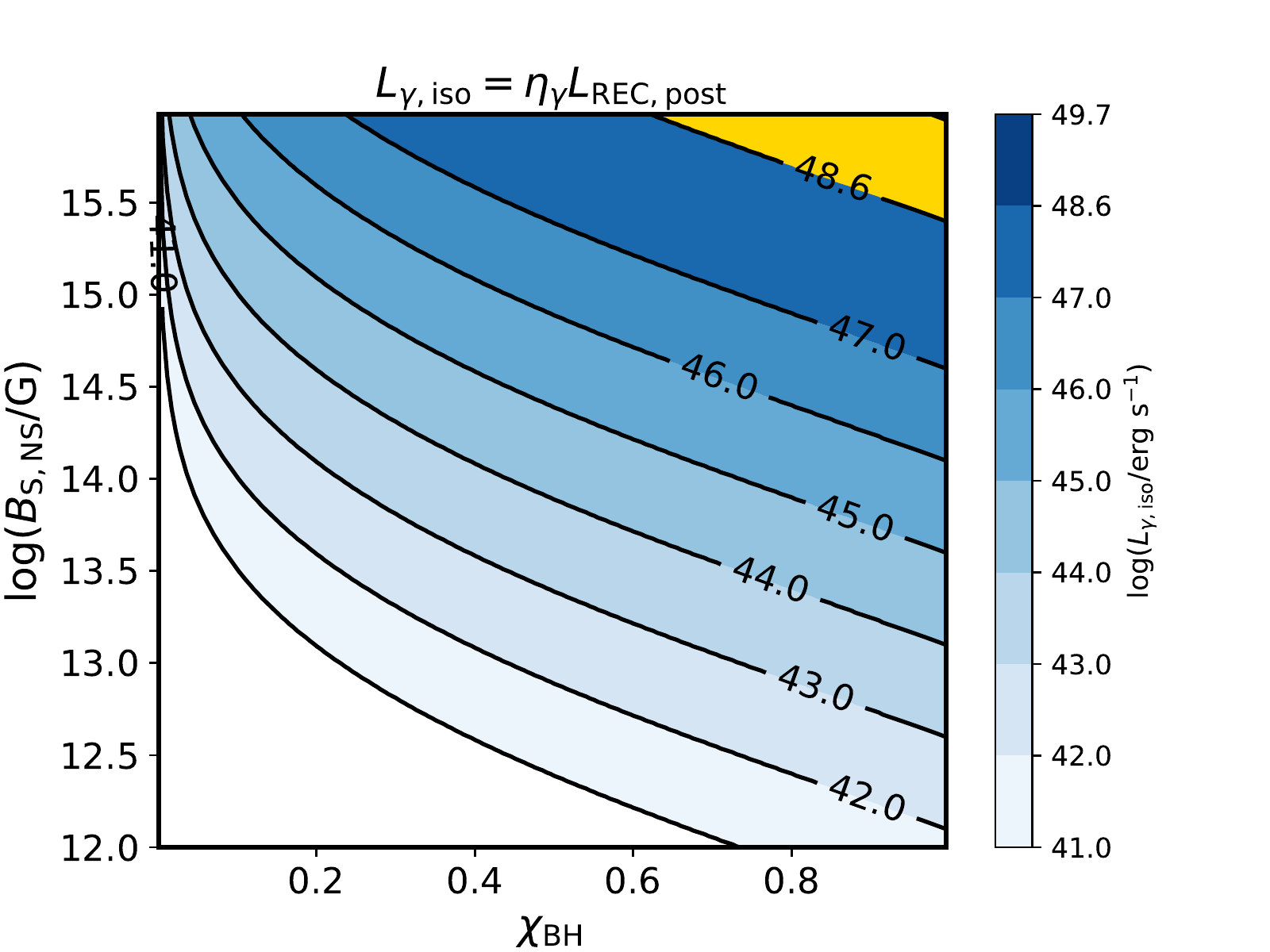}
\includegraphics[angle=0,width=0.5\textwidth]{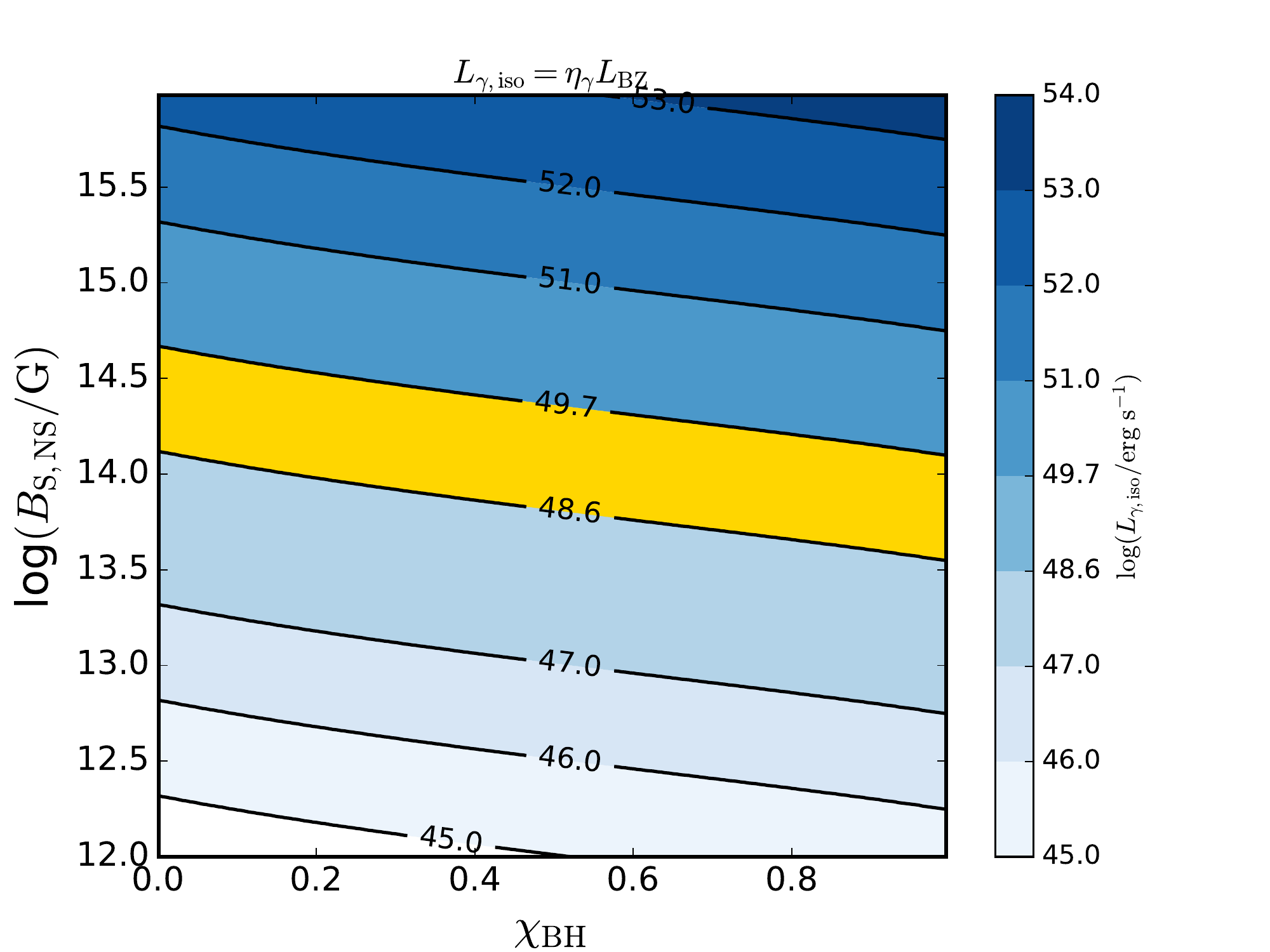}
\caption{The contours of isotropic gamma-ray radiation luminosities from various pre-merger and the post-merger mechanisms. From left to right and top to down the six panels denote the first magnetic dipole radiation (MDR,1), the second magnetic dipole radiation (MDR,2), the electric dipole radiation (EDR), the pre-merger magnetic reconnection (REC,pre),  the post-merger magnetic reconnection (REC,post), and the BZ mechanism (BZ). All the contours are plotted in the plane of the pre-merger BH spin $\chi_{\rm BH}$ and NS surface magnetic field strength $B_{\rm S,NS}$. The {\em yellow regions} represent the isotropic gamma-ray radiation luminosity ${\rm log}(L_{\gamma,\rm iso}/{\rm erg~s^{-1}})\sim48.6-49.7$ of the sub-threshold GRB GBM-190816. The radiative efficiency is adopted as $\eta_{\gamma}=1$. The mass ratio $q=5$}, the NS mass $M_{\rm NS}=1.4~M_{\odot}$ and radius $r_{\rm NS}=12~$km are adopted in the numerical calculations.
\label{fig:mechanisms}
\end{figure*}

This scenario involves a plunging BH--NS system in which the BH is immersed in the magnetic field of the NS and gains charge via the Wald mechanism \citep{wald74} in an initial electro-vacuum approximation. \cite{levin18} suggested that the BH can be charged stably to carry the Wald's charge quantity $Q_{\rm W}$ until it could transit from the electro-vacuum state to the force-free state thanks to abundant pair production induced by the strong electric field. At this point, the BH may reach the maximal Wald charge. In this scenario, there are four possible pre-merger mechanisms \citep[first and second magnetic dipole radiation, electric dipole radiation, and magnetic reconnection close to BH's equatorial plane;][]{Dai2019} and two possible post-merger mechanisms \citep[magnetic reconnection at polar regions and the BZ mechanism;][]{zhong19} to generate $\gamma$-ray emission. Following \cite{Dai2019} and \cite{zhong19}, we calculate (Figure \ref{fig:mechanisms}) that the sub-threshold GRB could be produced by the pre-merger magnetic reconnection or the post-merger BZ mechanism if the NS surface magnetic field satisfies $\log (B_{\rm S,NS}/{\rm G})>{13.5}$ or $\log (B_{\rm S,NS}/{\rm G})\sim13.5-14.6$, respectively, given the following conditions: the radiative efficiency $\eta_{\gamma}=1$, the mass ratio is $q=5$, the minimal separation between the BH and the NS is $a_{\rm min}=2GM_{\rm BH}/c^2+r_{\rm NS}$, the NS mass is $M_{\rm NS}=1.4~M_{\odot}$ and its radius is $r_{\rm NS}=12~$km. The following two points are worth mentioning in our calculation:
(1) We consider that the pre-merger magnetic reconnection in Equation (19) of \cite{Dai2019} should be the BH's magnetic field produced by the Wald charge $Q_{\rm W}$ rather than that of the NS. This is because the BH's magnetic field should be always lower than that of the NS, as pointed out in \cite{levin18}.
(2) For the post-merger magnetic reconnection and BZ mechanism, the parameters such as the BH's spin and mass and their derived parameters should be relevant to the final BH rather than the pre-merger BH in the binary system. However, they can be linked to those of the pre-merger BH through Equations (\ref{eq:x_BH,f}) with $M=M_{\rm BH}+M_{\rm NS}$. 

\subsection{The GW-GRB Delay Timescale}
\label{sec:timedelay}
	
The delay time between GBM-190816 and the putative gravitational wave event is about 1.57 s {in the observer frame and is about 1.43 s in the cosmological proper frame}. This is similar to the 1.70 s GW-GRB delay observed in GW170817 / GRB 170817A \citep{Abbott2017} which is also 1.68 s in the cosmological proper frame. In the literature, the origin of 1.7 s delay has been extensively discussed. The delay due to the effects of exotic physics is likely small \citep{Wei2017,Shoemaker2018,Burns2019}, and the main contribution is likely due to astrophysical processes \citep{zhang18,BZhang2019b}.

Following the convention introduced in \cite{BZhang2019b}, we discuss the three terms of the astrophysical GW-GRB delay timescale. Since the cCBC scenario is not favored, we limit ourselves to the hyperaccreting NS-BH merger scenario. 

(1) $\Delta t_{\rm jet}$: the delay time to launch a clean relativistic jet. In general, such a delay includes three parts for a hyperaccreting BH central engine, namely, the waiting time $\Delta t_{\rm wait}$ for a central object (BH) to form, the accretion time scale $\Delta t_{\rm acc}$, and the time $\Delta t_{\rm clean}$ for the jet to become clean. In our considered scenario for the event GBM-190816, since at least one BH already exists in the pre-merger system, $\Delta t_{\rm wait}$ should be 0. For a black hole engine, $\Delta t_{\rm clean} \sim 0$ and $\Delta t_{\rm acc}$ is typically $\sim 10$ ms. So $\Delta t_{\rm jet}$ is $\sim  0.01$ s.

(2) $\Delta t_{\rm bo}$: the delay time for the jet to break out from the surrounding medium. For an NS-BH progenitor, this time scale is typically $10-100$ ms. 

(3) $\Delta t_{\rm GRB}$: the delay time for the jet to reach the energy dissipation and GRB emission site. Such a delay is directly related to the emission radius, i.e., $t_{\rm GRB}=R/2c\Gamma^2$, where $\Gamma$ is the Lorentz factor of the eject and $c$ is the speed of light. In view that the first two terms are negligibly small for NS-BH mergers, the {cosmological-proper-frame 1.43 s delay} should be mainly defined by this term. The falling time scale of a burst is defined by the angular spreading time, which carries the same expression as $t_{\rm GRB}$, one would then expect that the true duration of GBM-190816 would be of the same order of the delay time scale (1.43 s). The observed $T_{90} \sim 0.1$ s is apparently much shorter than this. However, it is possible that the true burst is longer and the observed $T_{90}$ is simply the tip-of-iceberg of the true burst. The fact that the amplitude parameter $f$ is not very large allows such a possibility. 

The fact that the {1.43} s-delay in GBM-190816 is similar to the 1.68 s-delay in GW170817/GRB 170817A also sheds light on the origin of the delay in the latter system. Since GW170817/GRB 170817A is an NS-NS merger system, the final merger product is quite uncertain, which depends on the unknown neutron star equation of state \citep{ai2020}. If the merger product turns into a black hole before the GRB jet is launched, it is possible that there is a significant delay attributable to $\Delta t_{\rm jet}$ \citep[e.g.][]{Nakar2018}. However, this scenario has to introduce chance coincidence to explain the apparent consistency between the delay time and the duration of the burst. Alternatively, if the merger product does not collapse to a black hole before the jet is launched, then there is no immediate reason to suggest the existence of a significant $\Delta t_{\rm jet}$. The fact of a comparable delay time and duration then favors the possibility that $\Delta t_{\rm GRB}$ is the dominant contribution to the observed $\Delta t$ \citep{zhang18,BZhang2019b}.

Since for NS-BH mergers, the observed time delay should be mostly contributed by $\Delta t_{\rm GRB}$, the fact that the GBM-190816 has a comparable amount of the delay from its GW counterpart suggests that $\Delta t_{\rm GRB}$ itself can be this long. This indirectly suggests that the jet in GW170817/GRB 170817A was launched promptly without significant delay \citep{zhang18,BZhang2019b}. This conclusion is also supported by a recent independent study of \cite{Beniamini2020}.

\section{Conclusions}

In this paper, we performed a comprehensive study of the sub-threshold GRB GBM-190816 that is associated with a sub-threshold GW event. Based on publicly available information, we present the properties of the burst and discussed the physical implications of the data. Our key findings are the following:

(1) By studying the temporal and spectral properties of GBM-190816 and comparing them with those of other short GRBs, we confirm that GBM-190816 can be classified as a weak short GRB. 

(2) Based on the available information of the sub-threshold GW event, we were able to constrain the mass ratio of the binary as $q \sim q=2.26_{-0.12}^{+2.75}$.

(3) The association, if real, is mostly due to an NS-BH merger with tidal disruption. The constraints on the mass ratio $q$, BH spin, and viewing angle are derived based on the hyperaccretion BH central engine model and a Gaussian structured jet geometric model.

(4) We also discussed the scenarios of charged CBCs to produce the observed GRB. For the constant charge models, the required charge is much larger than what is expected, suggesting that these scenarios do not work unless contrived physical conditions are imposed. For the plunging NS-BH mergers with an increasing charge of the BH, the standard magnetic dipole radiation and electric dipole radiation components also cannot meet the observed luminosity unless extreme parameters (e.g. the pre-merger BH spin) are invoked. However, a GRB with the observed luminosity may be produced through the pre-merger magnetic reconnection or post-merger BZ mechanism for not-too-extreme parameters.

(5) By comparing the GW-GRB delay timescales between this event and GW170817/GRB 170817A, we conclude that the GW-GRB delay of these two cases is mostly contributed by the time scale for the jet to reach the energy dissipation radius where the observed $\gamma$-rays are emitted.

We note that our conclusions above are based on the assumption that the association between the GBM-190816 and the sub-threshold GW event is real. Further confirmation is needed by the more detailed joint analysis of the GW data and the GRB data by the LIGO/Virgo/Fermi team. In any case, the theoretical framework developed in this paper can be applied to this and other future CBC events with GRB associations, especially those originating from NS-BH mergers.

\acknowledgments
We thank Eric Burns for important information and the anonymous referee for helpful suggestions. 
BBZ acknowledges support from a national program for young scholars in China. This work is supported by National Key Research and Development Programs of China (2018YFA0404204, 2017YFA0402600) and The National Natural Science Foundation of China (Grant Nos. 11833003, 11722324, 11633001, 11690024,11573014). JSW is supported by China Postdoctoral Science Foundation. This work is also supported by NSFC 11922301 (HJL). We acknowledge the use of public data from the Fermi Science Support Center (FSSC). This research has made use of data, software and/or web tools obtained from the Gravitational Wave Open Science Center (https://www.gw-openscience.org), a service of LIGO Laboratory, the LIGO Scientific Collaboration and the Virgo Collaboration.

\begin{table*}
		\begin{center}
			\caption{Spectral properties of GBM-190816 using the Fit of cut-off power law model }
			\label{tab:cplpara}
			\begin{tabular}{cc|cccccccccc}
				\hline
				\hline

				Time Interval & & & & CPL & & \\
				
				\cline{1-7}
				
				$t_1$ &$t_2$&$\Gamma_{\rm ph}$ &$E_{\rm p}$ &$logNorm$ & PGSTAT/dof \\
				\hline

				0.032& 0.143&$-0.92_{-0.58}^{+0.32}$ & $94.84_{-17.94}^{+114.64}$&$0.53_{-0.41}^{+0.72}$ & 130.1/227 \\
				
				\hline
				\hline
			\end{tabular}
		\end{center}
\end{table*}

\begin{table*}
		\begin{center}
			\caption{Observational Properties and Derived Constraints of GBM-190816.}
			\label{tab:para}
			\begin{tabular}{ll}
				\hline
				\hline
				Observed Properties & \\
				
				\hline

				T$_{\rm 90}$ (s) & $0.112_{-0.085}^{+0.185}$ \\

				Peak energy $E_{\rm p}$ (keV) & $94.84_{-17.94}^{+114.64}$ \\
				Total fluence($\rm erg\,cm^{-2}$) & $7.38_{-2.51}^{+6.35} \times 10^{-8}$ \\
				Distance (Mpc) & $428 +/- 143$ \\
 Isotropic energy $E_{\gamma,\rm iso}$ (erg) & $1.65_{-1.16}^{+3.81} \times 10^{48} $ \\
 Luminosity $L_{\gamma,\rm iso}$ ($\rm erg\,s^{-1}$) & $1.47_{-1.04}^{+3.40} \times 10^{49} $ \\
 \textit{f} parameter& $2.58 +/- 0.37$\\
 				\hline
				Assumed Parameters& \\
				\hline
				Jet core angle $ \theta_{\rm c,j}$ & assumed 5$^{\circ}$ (16$^{\circ}$)\\
				Viewing angle $\theta_{\rm v}$ & $10^{\circ}-19^{\circ}$ ($18^{\circ}-24^{\circ}$)\\
				$\Gamma_{\rm c}$ & assumed 100\\
				$m_2$ ($M_{\odot}$) & assumed 1.4 (for NS-BH system)\\
				& assumed 2.8 (for BH-BH system)\\
				\hline

				Derived Constrains& \\
				
				\hline
				
				$q$ from GRB & varies\\
				$q$ from GW & $2.26_{-0.12}^{+2.75}$ \\
				$m_1$ ($M_{\odot}$) & varies\\
				GW-GRB Time Delay (s)&1.57\\
				Charge of BH (e.s.u.)& $2.162_{-0.002}^{+0.302} \times 10^{26} $ (for NS-BH system)\\
				& $2.114_{-0.042}^{+1.131}\times 10^{26}$ (for BH-BH system)\\
				
				\hline
				\hline
			\end{tabular}
		\end{center}
\end{table*}

\newpage

\newpage
 
\newpage

\begin{thebibliography}{}
\bibitem[Abbott et al.(2016)]{Abbott2016} Abbott, B.~P., Abbott, R., Abbott, T.~D., et al.\ 2016, \prl, 116, 061102
\bibitem[Abbott et al.(2017)]{Abbott2017} Abbott, B.~P., Abbott, R., Abbott, T.~D., et al.\ 2017, \prl, 119, 161101
\bibitem[Abbott et al.(2019)]{Abbott2019} Abbott, B.~P., Abbott, R., Abbott, T.~D., et al.\ 2019, Physical Review X, 9, 031040
\bibitem[Ai et al.(2020)]{ai2020} Ai, S., Gao, H., \& Zhang, B., \ 2020, \apj, 893, 146
\bibitem[Allen et al.(2012)]{Allen2012} Allen, B., Anderson, W.~G., Brady, P.~R., et al.\ 2012, \prd, 85, 122006
\bibitem[Amati et al.(2002)]{Amati2002} Amati, L., Frontera, F., Tavani, M., et al.\ 2002, \aap, 390, 81
\bibitem[Barbieri et al.(2019)]{barb19} Barbieri, C., Salafia, O.~S., Perego, A., et al.\ 2019, arXiv e-prints, arXiv:1908.08822
\bibitem[Bardeen et al.(1972)]{bard72} Bardeen, J.~M., Press, W.~H., \& Teukolsky, S.~A.\ 1972, \apj, 178, 347
\bibitem[Belczynski et al.(2002)]{Belczynski2002} Belczynski, K., Kalogera, V., \& Bulik, T.\ 2002, \apj, 572, 407
\bibitem[Beniamini et al.(2019)]{Beniamini2019} Beniamini, P., Petropoulou, M., Barniol Duran, R., et al.\ 2019, \mnras, 483, 840
\bibitem[Beniamini et al.(2020)]{Beniamini2020} Beniamini, P., Barniol Duran, R., Petropoulou, M., et al.\ 2020, arXiv e-prints, arXiv:2001.00950
\bibitem[Blandford \& Znajek(1977)]{bland77} Blandford, R.~D., \& Znajek, R.~L.\ 1977, \mnras, 179, 433
\bibitem[Buonanno et al.(2008)]{buo08} Buonanno, A., Kidder, L.~E., \& Lehner, L.\ 2008, \prd, 77, 026004
\bibitem[Burns(2019)]{Burns2019} Burns, E.\ 2019, arXiv e-prints, arXiv:1909.06085
\bibitem[Callister et al.(2017)]{Callister2017} Callister, T.~A., Kanner, J.~B., Massinger, T.~J., et al.\ 2017, Classical and Quantum Gravity, 34, 155007
\bibitem[Connaughton et al.(2016)]{Connaughton16} Connaughton, V., Burns, E., Goldstein, A., et al.\ 2016, \apjl, 826, L6
\bibitem[Cutler, \& Flanagan(1994)]{Cutler1994} Cutler, C., \& Flanagan, {\'E}. E.\ 1994, \prd, 49, 2658
\bibitem[D'Orazio et al.(2016)]{do16} D'Orazio, D.~J., Levin, J., Murray, N.~W., et al.\ 2016, \prd, 94, 23001
\bibitem[Dai(2019)]{Dai2019} Dai, Z.~G.\ 2019, \apjl, 873, L13
\bibitem[Deng et al.(2018)]{Deng2018} Deng, C.-M., Cai, Y., Wu, X.-F., et al.\ 2018, \prd, 98, 123016
\bibitem[Fong et al.(2015)]{fong15} Fong, W., Berger, E., Margutti, R., et al.\ 2015, \apj, 815, 102
\bibitem[Foucart et al.(2018)]{fou18} Foucart, F., Hinderer, T., \& Nissanke, S.\ 2018, \prd, 98, 081501
\bibitem[Foucart et al.(2019)]{fou19} Foucart, F., Duez, M.~D., Kidder, L.~E., et al.\ 2019, \prd, 99, 103025
\bibitem[Gao et al.(2020)]{gao19} Gao, H., Ai, S.-K., Cao, Z.-J., et al.\ 2020, Frontiers of Physics, 15,  24603
\bibitem[Ghirlanda et al.(2019)]{ghir19} Ghirlanda, G., Salafia, O.~S., Paragi, Z., et al.\ 2019, Science, 363, 968
\bibitem[Goldstein et al.(2019)]{Goldstein2019a} Goldstein, A., Hamburg, R., Wood, J., et al.\ 2019, arXiv e-prints, arXiv:1903.12597
\bibitem[Goldstein et al.(2017)]{Goldstein2017} Goldstein, A., Veres, P., Burns, E., et al.\ 2017, \apjl, 848, L14
\bibitem[Hawley et al.(2015)]{haw15} Hawley, J.~F., Fendt, C., Hardcastle, M., et al.\ 2015, \ssr, 191, 441
\bibitem[Kawaguchi et al.(2016)]{kaw16} Kawaguchi, K., Kyutoku, K., Shibata, M., et al.\ 2016, \apj, 825, 52
\bibitem[Lattimer \& Prakash(2001)]{latt01} Lattimer, J.~M., \& Prakash, M.\ 2001, \apj, 550, 426
\bibitem[Lazzati et al.(2018)]{Lazzati2018} Lazzati, D., Perna, R., Morsony, B.~J., et al.\ 2018, \prl, 120, 241103
\bibitem[Levin et al.(2018)]{levin18} Levin, J., D'Orazio, D.~J., \& Garcia-Saenz, S.\ 2018, \prd, 98, 123002
\bibitem[LIGO/Virgo/Fermi Collaboration(2019a)]{Goldstein2019} LIGO/Virgo/Fermi Collaboration.\ 2019, GCN, 25406. 
\bibitem[LIGO/Virgo/Fermi Collaboration(2019b)]{Shawhan2019} LIGO/Virgo/Fermi Collaboration.\ 2019, GCN, 25465.
\bibitem[L{\"u} et al.(2014)]{lv2014} L{\"u}, H.-J., Zhang, B., Liang, E.-W., et al.\ 2014, \mnras, 442, 1922
\bibitem[Lyman et al.(2018)]{Lyman2018} Lyman, J.~D., Lamb, G.~P., Levan, A.~J., et al.\ 2018, Nature Astronomy, 2, 751
\bibitem[McWilliams \& Levin(2011)]{mcw11} McWilliams, S.~T., \& Levin, J.\ 2011, \apj, 742, 90
\bibitem[Meegan et al.(2009)]{Meegan2009} Meegan, C., Lichti, G., Bhat, P.~N., et al.\ 2009, \apj, 702, 791.
\bibitem[Michel(1982)]{Michel1982} Michel, F.~C.\ 1982, Reviews of Modern Physics, 54, 1
\bibitem[Nakar, \& Piran(2018)]{Nakar2018} Nakar, E., \& Piran, T.\ 2018, \mnras, 478, 407
\bibitem[Nitz et al.(2020)]{Nitz2020} Nitz, A.~H., Dent, T., Davies, G.~S., et al.\ 2020, \apj, 891, 123
\bibitem[Pan \& Yang(2019)]{pan19} Pan, Z., \& Yang, H.\ 2019, \prd, 100, 043025
\bibitem[Pannarale(2013)]{pann13} Pannarale, F.\ 2013, \prd, 88, 104025
\bibitem[Poisson, \& Will(1995)]{Poisson1995} Poisson, E., \& Will, C.~M.\ 1995, \prd, 52, 848
\bibitem[Popham et al.(1999)]{pop99} Popham, R., Woosley, S.~E., \& Fryer, C.\ 1999, \apj, 518, 356
\bibitem[Ryan et al.(2019)]{Ryan2019} Ryan, G., van Eerten, H., Piro, L., et al.\ 2019, arXiv e-prints, arXiv:1909.11691
\bibitem[Sathyaprakash, \& Dhurandhar(1991)]{Sathyaprakash1991} Sathyaprakash, B.~S., \& Dhurandhar, S.~V.\ 1991, \prd, 44, 3819
\bibitem[Savchenko et al.(2017)]{Savchenko2017} Savchenko, V., Ferrigno, C., Kuulkers, E., et al.\ 2017, \apjl, 848, L15
\bibitem[Scargle et al.(2013)]{Scargle2013} Scargle, J.~D., Norris, J.~P., Jackson, B., et al.\ 2013, \apj, 764, 167
\bibitem[Shibata et al.(2009)]{Shibata2009} Shibata, M., Kyutoku, K., Yamamoto, T., et al.\ 2009, \prd, 79, 044030
\bibitem[Shoemaker, \& Murase(2018)]{Shoemaker2018} Shoemaker, I.~M., \& Murase, K.\ 2018, \prd, 97, 083013
\bibitem[Tchekhovskoy et al.(2010)]{tch10} Tchekhovskoy, A., Narayan, R., \& McKinney, J.~C.\ 2010, \apj, 711, 50
\bibitem[The LIGO Scientific Collaboration et al.(2020)]{Abbott2020} The LIGO Scientific Collaboration, the Virgo Collaboration, Abbott, B.~P., et al.\ 2020, arXiv e-prints, arXiv:2001.01761
\bibitem[Troja et al.(2019)]{Troja2019} Troja, E., van Eerten, H., Ryan, G., et al.\ 2019, \mnras, 489, 1919
\bibitem[Tsang et al.(2012)]{tsang12} Tsang, D., Read, J.~S., Hinderer, T., et al.\ 2012, \prl, 108, 011102

\bibitem[Veitch et al.(2015)]{Veitch2015} Veitch, J., Raymond, V., Farr, B., et al.\ 2015, \prd, 91, 042003
\bibitem[Wald(1974)]{wald74} Wald, R.~M.\ 1974, \prd, 10, 1680
\bibitem[Wang et al.(2019)]{Wang2019} Wang, J. S., et al. 2019, in preparation
\bibitem[Wei et al.(2017)]{Wei2017} Wei, J.-J., Zhang, B.-B., Wu, X.-F., et al.\ 2017, \jcap, 2017, 035
\bibitem[Yagi \& Yunes(2017)]{yagi17} Yagi, K., \& Yunes, N.\ 2017, \physrep, 681, 1

\bibitem[Zhang(2016)]{BZhang2016} Zhang, B.\ 2016, \apjl, 827, L31
\bibitem[Zhang(2019a)]{BZhang2019a} Zhang, B.\ 2019a, \apjl, 873, L9
\bibitem[Zhang(2019b)]{BZhang2019b} Zhang, B.\ 2019b, Frontiers of Physics, 14, 64402
\bibitem[Zhang \& M{\'e}sz{\'a}ros(2002)]{Zhang2002} Zhang, B., \& M{\'e}sz{\'a}ros, P.\ 2002, \apj, 571, 876

\bibitem[Zhang et al.(2009)]{BZhang2009} Zhang, B., Zhang, B.-B., Virgili, F.~J., et al.\ 2009, \apj, 703, 1696

\bibitem[Zhang et al.(2016)]{BBZhang2016} Zhang, B.-B., Uhm, Z.~L., Connaughton, V., et al.\ 2016, \apj, 816, 72
\bibitem[Zhang et al.(2018a)]{BBZhang2018} Zhang, B.-B., Zhang, B., Castro-Tirado, A.~J., et al.\ 2018a, Nature Astronomy, 2, 69.
\bibitem[Zhang et al.(2011)]{BBZhang2011} Zhang, B.-B., Zhang, B., Liang, E.-W., et al.\ 2011, \apj, 730, 141
\bibitem[Zhang et al.(2018b)]{zhang18} Zhang, B.-B., Zhang, B., Sun, H., et al.\ 2018b, Nature Communications, 9, 447
\bibitem[Zhong et al.(2019)]{zhong19} Zhong, S.-Q., Dai, Z.-G., \& Deng, C.-M.\ 2019, \apjl, 883, L19



\clearpage
\end{thebibliography}
\end{document}